\documentclass[]{pasj01}
\usepackage{lscape}
\draft 

\begin{document} 

\title{Fluctuation of the background sky in the Hubble Extremely Deep Field (XDF) and its origin}

\author{Toshio \textsc{Matsumoto}\altaffilmark{1}} \author{Kohji \textsc{Tsumura}\altaffilmark{2,3}}

\altaffiltext{1}{Department of Space Astronomy and Astrophysics, Institute of Space and 
Astronautical Science, Japan Aerospace Exploration Agency, Sagamihara, Kanagawa 252-5210, Japan}
\altaffiltext{2}{Department of Natural Science, Faculty of Knowledge Engineering, Tokyo City University,
Setagaya, Tokyo, 158-8557, Japan}
\altaffiltext{3}{Frontier Research Institute for Interdisciplinary Science,
Tohoku University, Sendai, Miyagi 980-8578, Japan}

\email{matsumo@ir.isas.jaxa.jp}


\maketitle

\begin{abstract}
We performed a fluctuation analysis of the Hubble Extremely Deep Field (XDF) at four optical
wavelength bands and found large fluctuations
that are significantly brighter than those expected for ordinary galaxies. Good cross-correlations with flat spectra are
found down to 0.2 arcsec, indicating the existence of a spatial structure even at the 0.2 arcsec scale.
The detected auto and cross-correlations provide a lower limit of 24 nW m$^{-2}$ sr$^{-1}$ for the absolute sky brightness  
 at {\bf $700 \sim 900$} nm, which is consistent with previous observations.

We searched for candidate objects to explain the detected large fluctuation using the catalog of the Hubble Ultra Deep Field (UDF), 
and found that the surface number density of faint compact objects (FCOs) rapidly increases toward 
the faint end. Radial profiles of FCOs are indistinguishable 
from the PSF, and the effective radius based on de Vaucouleur's law is estimated to be
smaller than 0.02 arcsec. 
The SEDs of FCOs follow a power law at optical wavelengths, but show greater emission and structure at {\bf $\lambda > 1 \mu$m}.
Assuming that the FCOs are the cause of the excess brightness and fluctuations, the 
faint magnitude limit is 34.9 mag for the F775W band,
and the surface number density reaches  {\bf $2.6 \times 10^{3}$ (arcsec)$^{-2}$}.

Recent {\bf $\gamma$} ray observations require that the redshift of FCOs
must be less than 0.1, if FCOs are the origin of the excess optical and infrared background. Assuming that FCOs consist of missing baryons,
the mass and luminosity of a single FCO range from $10^{2}$ to 1$0^{3}$ solar units, and mass-to-luminosity ratio is significantly
lower than 1.0 solar unit. The maximum effective radius of an FCO is 4.7 pc.
These results and the good correlation between near-infrared and X-ray background
indicate that FCOs could be powered by 
the gravitational energy associated with black holes.
\end{abstract}

\section{Introduction}
The extragalactic background light (EBL) is thought to represent the energy density of the universe, and has been
observed over a wide range of wavelengths. In the optical and near-infrared regions, EBL observations 
have been performed and excess brightness over the integrated light of known celestial sources has been detected based on the data of 
HST (\citet{Bernstein07}, \citet{Kawara17}), DIRBE/COBE (\citet{Hauser98}, \citet{Wright98}, \citet{Wright00},
\citet{Gorjian00}, \citet{Cambrecy01}, \citet{Levenson07}, \citet{Sano15}, \citet{Sano16}), 
NIRS/IRTS  (\citet{Matsumoto05}, \citet{Matsumoto15}), AKARI \citep{Tsumura13}, and CIBER \citep{Matsuura17} etc.
The detection of EBL is not simple due to the bright foreground emission known as the zodiacal light. 
However, the zodiacal light is so sufficiently smooth \citep{Pyo12} for the detection of the EBL fluctuation that
the measurements have been conducted with Spitzer (\citet{Kashlinsky05}, \citet{Kashlinsky07}, \citet{Kashlinsky12}),
AKARI (\citet{Matsumoto11}, \citet{Seo15}), HST (\citet{Thompson07}, \citet{M-W15}, \citet{Donnerstein15}), CIBER  \citep{Zemcov14}, and
IRTS  \citep{Kim19}. Researchers have found excess fluctuation that cannot be explained by ordinary galaxies \citep{Helgason12}.

The origin of the excess brightness and fluctuation has been examined by numerous researchers. First stars are an attractive candidate, however, 
recent evolution models based on the observations of high-redshift galaxies indicate that the brightness and fluctuation caused by
first stars are much lower than the observed excesses (\citet{Cooray12a}, \citet{Yue13a}).  As an origin of the excess fluctuation,
\citet{Cooray12b} proposed the intra-halo light (IHL) defined as the emission from stars expelled into the intergalactic space due to 
collisions and/or merging of galaxies at low redshifts. IHL is advantageous in energetics problems, however, 
no observational evidence of IHL has yet been reported. Furthermore, there are difficulties in invoking IHL to
explain large fluctuations at the degree scale \citep{Kim19} and the excess EBL. 
\citet{Yue13b} explained the fluctuation observed with Spitzer using direct 
collapsed black holes (DCBHs) at high redshifts. DCBHs are favorable for explaining the cross-correlation between the near-infrared and
X-ray background (\citet{Cappelluti13}, \citet{M-W15}, \citet{Cappelluti17}), however, difficulties remain for explaining 
the observed fluctuation at $\lambda < 1\mu$m 
 (\citet{M-W15}, \citet{Zemcov14}).

At present, there is no known source that can explain all of the EBL observations. To obtain new observational evidence,
we performed a fluctuation analysis for the deepest images, Hubble Extreme Deep Field (XDF) \citep{Illingworth13} with 
four filter bands (F435W, F606W, F775W, and F850LP). Our approach is very similar to the fluctuation
analysis of the NICMOS Ultra Deep Field (NUDF) by \citet{Thompson07} and \citet{Donnerstein15} at 1.0 and 1.6  $\mu$ m,
and  of the Cosmic Assembly Near-Infrared Deep Extragalactic Legacy Survey (CANDELS) field 
at optical and near-infrared wavelengths \citep{M-W15}.
The detection limits of the XDF sources are $\sim$ 30 mag (AB) \citep{Illingworth13}, 
which are two or three magnitude deeper than those in previous analyses. This value implies that the contribution of the 
ordinary faint galaxies to the surface brightness in the XDF blank sky is very low, which may enable one to observe the nature of unknown
objects. Cross-correlations between wavelength bands are also analyzed to determine the spatial structure of the fluctuation.

We have attempted to identify candidates of the unknown sources in the XDF, and discovered the faint compact objects (FCOs). 
Assuming that FCOs are responsible for the excess brightness and fluctuation in the sky, we examined the properties
of FCOs based on the Ultra Deep Field (UDF) \citep{Beckwith06} and 
ultraviolet UDF (UVUDF)  \citep{Rafelski15} catalogs.

This paper is organized as follows. In section 2, we present a fluctuation analysis of the XDF data, and 
the absolute sky brightness estimated from the detected fluctuation. In section 3, we examine the properties of
FCOs as the origin of the detected EBL excess. In section 4, we discuss the possible origin of FCOs, and a summary is given in
section 5.

\section{Fluctuation analysis}
XDF images are the deepest images observed with ACS/WFC aboard the HST \citep{Illingworth13}. 
The field of the optical bands is 11 arcmin$^{2}$ 
with images obtained from five filter bands. Their filter names and central bandwidths are F435W (432.3nm), 
F606W (591.2 nm), F775W (769.9 nm), F814W (818.6 nm),
 and F850LP (906.0 nm). We did not analyze the F814W image, because the integration time for this filter is significantly 
 lower than that of other bands, 
 and additionally no catalog is available for this band. We used XDF images from a publicly available 
 site\footnote{https://archive.stsci.edu/prepds/xdf/}, where images of two 
 pixel resolution levels (30 and 60 mas) are available. We used images with a resolution of 60 mas 
 for the fluctuation analysis because these images
have better sensitivity for diffuse extended sources.

The integration times of the XDF images are not uniform; instead, the times vary for
different locations. To achieve an accurate analysis,
we extracted the central region where the integration times are the longest and highly uniform (see Fig. 3 in \citet{Illingworth13}). 
The left panel of Figure 1 shows an extracted image for F775W.

A mask of the XDF images for fluctuation analysis was created by combining images for the four optical bands
 (F435W, F606W, F775W, and F850LP). The mask was created in four steps.
First, we masked all pixels that are brighter than  2$\sigma$, where $\sigma$ is the dispersion of the brightness distribution of the image. 
Almost all sources are masked in this procedure, but pixels surrounding bright objects are not perfectly masked because 
the diffuse extended wing around a bright object remains.
Thus, we treated adjacent 3 $\times$ 3 pixels of each pixel (including the original pixel) for masking to discriminate extended features.
In the second step, we additionally masked all 9 pixels when more than 4 pixels are brighter than 1$\sigma$. 
In the third step, we again masked all 9 pixels when more than 6 pixels are already masked. 
As the forth step, we repeated the third process to expand the mask by one more layer.
Finally, 53.3\% of the pixels remained after this masking procedure, and the extended wings of the 
objects were well masked (Figure 18).
The obtained mask was  applied to the four XDF optical images. The right panel of Figure 1 shows a masked image for the F775W band.
The brightness distributions of the masked images are very close to a Gaussian distribution 
(Figure 19). We  defined the limiting magnitudes
as 3$\sigma$ in the statistics for the average of  a 2 $\times$ 2 pixel square, 
which corresponds to 33.17, 33.62, 33.36, and 32.80 in AB magnitude for F435W, F606W, F775W, and F850LP, respectively.
Supplementary information on the masking is given in Appendix 1. 

\begin{figure}
 \begin{center}
  \includegraphics[width=16cm]{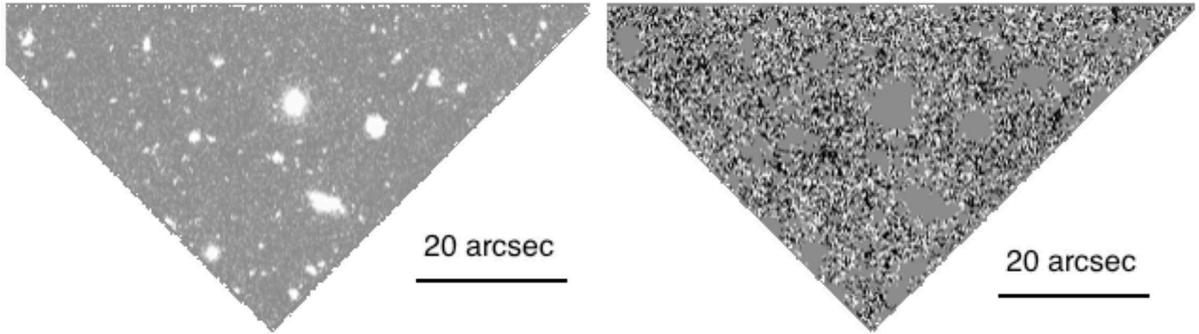} 
\end{center}
\caption{Extracted image for F775W. The left and right panels show images before and after 
masking, respectively. We changed the contrasts of the two images to clearly display the structure. }
 \label{fig1}
\end{figure}

For the masked images, we performed a two dimensional Fourier transformation and obtained power spectra, $P_{2}(q)$,
and fluctuations, \( [q^{2}P_{2}/(2\pi)]^{1/2} \), as a function of wavenumber, $q$, 
following previously reported analysis \citep{Thompson07}.

Since the fluctuation is artificially reduced at large angles when the number of the masked pixels is large,
we applied a simple correction for the masking effect. We created 100 simulation images with the same
fluctuation spectrum as the XDF F606W image, and estimated a recovery factor, defined as the ratio of the fluctuation levels  
before and after masking at a specific wave number, $q$. 
The recovery factor shows a strong dependence on the angular scale and amounts to $\sim1.5$ 
at the largest angle (Figure 19).
We applied the recovery factors to the fluctuation spectra, but the correction did not substantially influence the forthcoming analysis. 

Figure 2 shows the obtained auto fluctuations. Considering that the PSF is $\sim$ 0.1 arcsec, we plotted the data for angles larger than
0.2 arcsec. Shot noise at small angles and flat spectra at 10 - 30 arcsec are commonly present in these plots.  

\begin{figure}
 \begin{center}
  \includegraphics[width=16cm]{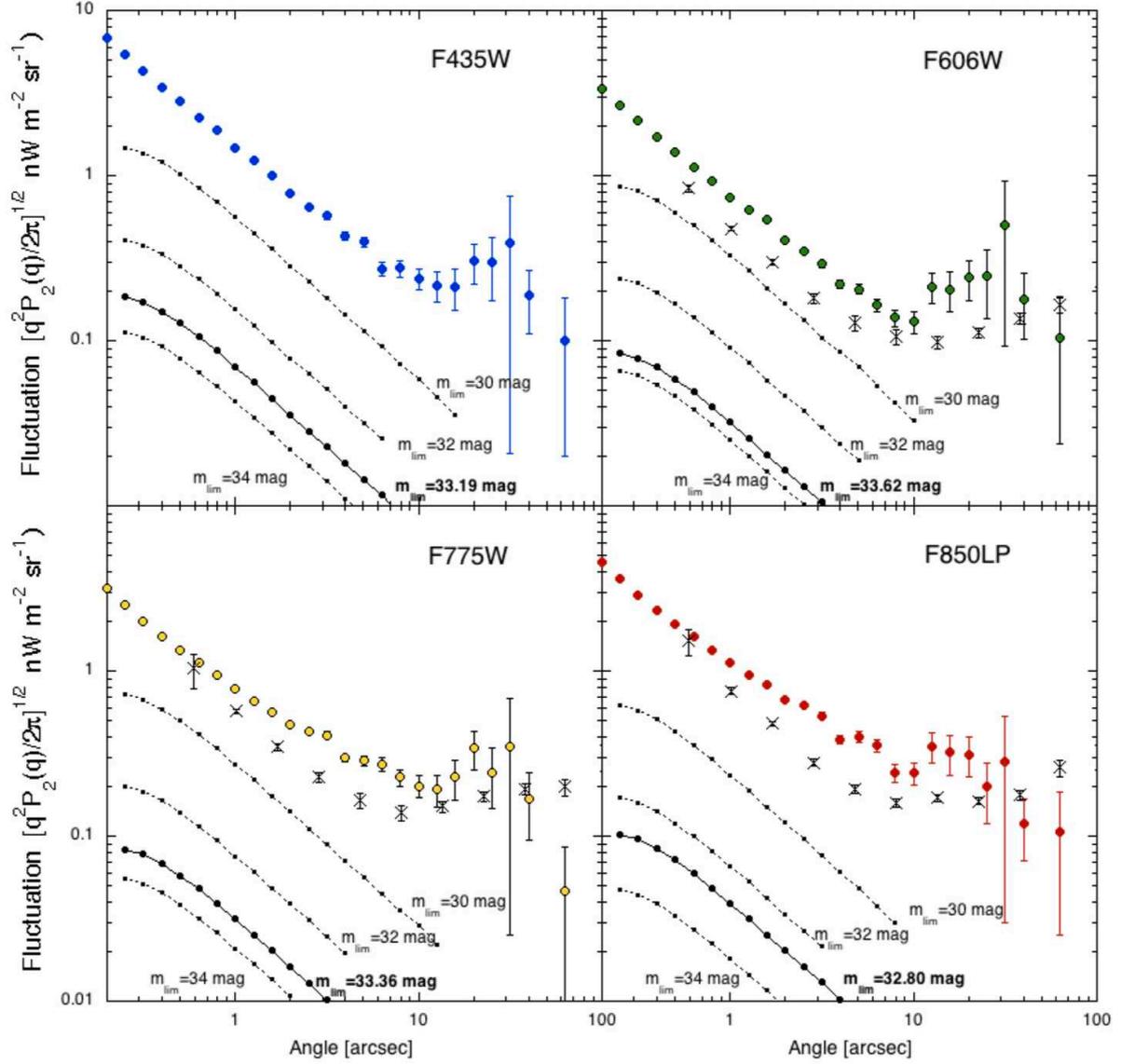} 
\end{center}
\caption{Auto-correlations of the masked XDF images for four optical bands. 
The crosses denote data obtained for the CANDELS field
 \citep{M-W15}. The dotted lines represent simulated fluctuations for ordinary galaxies 
with three different limiting magnitudes, m$_{lim}$. The solid lines indicate the same result obtained for
limiting magnitudes in our XDF fluctuation analysis. }
  \label{fig2}
\end{figure}

In Figure 2, we compared the observed fluctuations with those expected for ordinary galaxies, adopting the surface number
density and magnitude relations described in the next section (Fig. 7). 
The black dotted lines indicate fluctuations obtained via simulation analysis assuming a random distribution of 
ordinary galaxies that are fainter than limiting magnitudes of 30, 32, and 34 mag. 
The black solid lines correspond to the limiting magnitudes of the XDF masked images. 

We estimated the contribution of the clustering of faint galaxies based on the model by \citet{Helgason12} who examined the 
fluctuation due to one halo and two halo terms. The fluctuation of one halo term shows a similar feature (shot noise) to 
that of the random distribution, but is a few times lower than that of the random
distribution. The profile of two halo term has a peak at $\sim$1500 arcsec
and does not depend on the wavelengths. Furthermore, the absolute value at the peak is proportional to the 
integrated light of galaxies (ILG) 
fainter than the limiting magnitude. The ILG in our analysis ($\sim$ 0.06 nW m$^{-2}$ sr$^{-1}$) provides 
the peak fluctuations of two halo term to be   $\sim 2 \times 10^{-3}$ nW m$^{-2}$ sr$^{-1}$, which is negligible compared to the detected fluctuations.

Figure 2 and above considerations clearly show that the detected fluctuations are
significantly larger than those expected for ordinary galaxies, indicating the presence of unknown celestial objects. 
  
In Figure 2, fluctuations analyzed for imaging data from the CANDELS field are plotted as crosses \citep{M-W15}. Fluctuations 
for the F435W were not
analyzed for the CANDELS field. The fluctuation spectra for the CANDELS field are basically consistent with our XDF fluctuations,
but the absolute values are slightly lower than those of the XDF images. This  
difference is most likely caused by variations in image quality. The observation area
of the CANDELS field is much wider than that of the XDF, whereas the limiting magnitudes of the CANDELS survey are a few magnitude
lower than that of the XDF, although the limiting magnitudes are not explicitly reported by \citet{M-W15}. 
Furthermore, the number of overlapping images in the CANDELS analysis is not uniform; in the most extreme case,
there is ten-fold difference. These differences make it difficult to qualitatively compare the shot noise of the 
CANDELS field with the present result.

The contribution of the foreground signal is negligible. The fluctuation of the zodiacal light is less than 0.03\% \citep{Pyo12}
corresponding to $\sim$ 0.1 nW m$^{-2}$ sr$^{-1}$. Furthermore, the overlapping of a large number of images observed during different
seasons substantially reduces the fluctuation, since fluctuation of the zodiacal light is not fixed on the sky. Another foreground emission, 
diffuse Galactic light (DGL), shows a fluctuation proportional to the angular scale \citep{M-W15}, thus,
DGL fluctuation is negligible at the small angular scales utilized in this study. 

We attempted to perform a fluctuation analysis of the infrared XDF images, however, the absolute values of their fluctuations are
much lower than those of NUDF, as reported by  \citet{Thompson07} and CANDELS, as reported by \citet{M-W15}. 
This difference is most likely 
due to the strong spatial filters applied to correct for the effect of earthshine in the infrared XDF images \citep{Illingworth13}. 
Therefore, we do not include a fluctuation analysis of the infrared images in this paper.

Figure 3 shows the average fluctuations measured in the flat region ($\sim$20 arcsec)
as well as those of NUDF \citep{Thompson07} and CANDELS \citep{M-W15} at infrared wavelengths. 
Figure 3 shows that the fluctuations in the infrared bands are larger than those in the optical bands.

\begin{figure}
 \begin{center}
  \includegraphics[width=12cm]{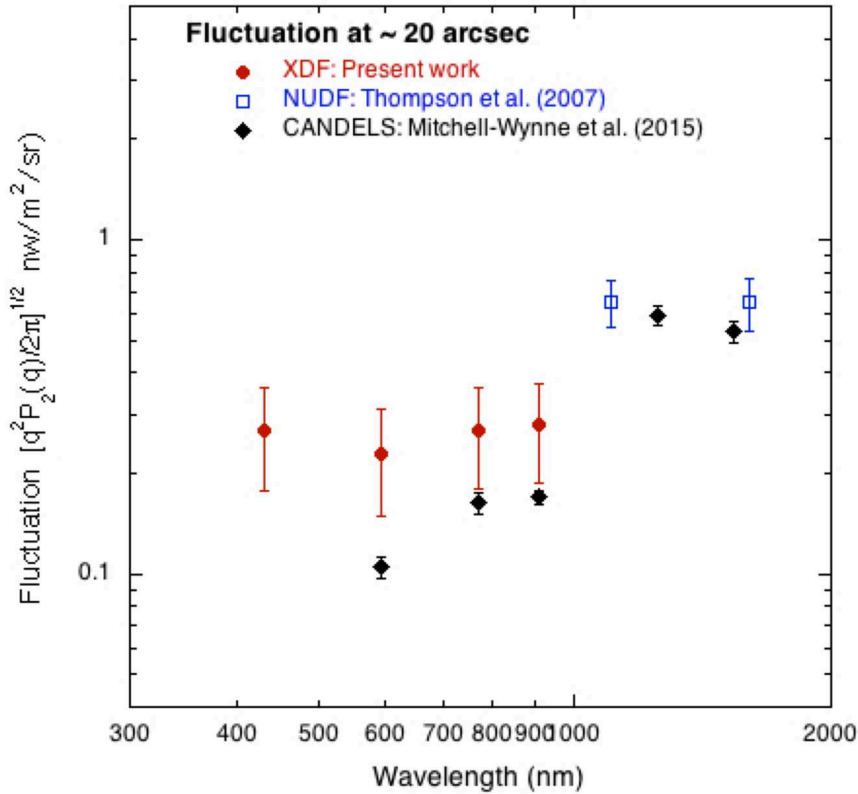} 
\end{center}
\caption{Dependence of the average fluctuations at $\sim 20$ arcsec on the wavelength (red filled circles). Fluctuations 
for the NUDF \citep{Thompson07} and CANDELS \citep{M-W15} fields are also plotted as open squares and 
solid diamonds, respectively.}
  \label{fig3}
\end{figure}

Figure 4 displays the cross-correlations between wavelength bands. Shot noise appears for the cross-correlations 
with F435W but is not observed for the cross-correlations with the three other bands, indicating
that the auto-correlation shot noises at small angular scales can be ascribed to photon noise and/or 
readout noise. The remaining shot noise for the cross-correlations with the F435W band may be due to faint ordinary galaxies, 
which will be examined in the next section.

In Figure 4, cross-correlations for the CANDELS field are plotted \citep{M-W15}. 
The basic features are similar to those of the XDF, however, shot noise components remain even for the three longer 
wavelength bands. This result can be ascribed to shot noises arising from the ordinary galaxies since the limiting magnitudes
in the CANDELS study are much lower than those in the present work.

\begin{figure}
 \begin{center}
  \includegraphics[width=16cm]{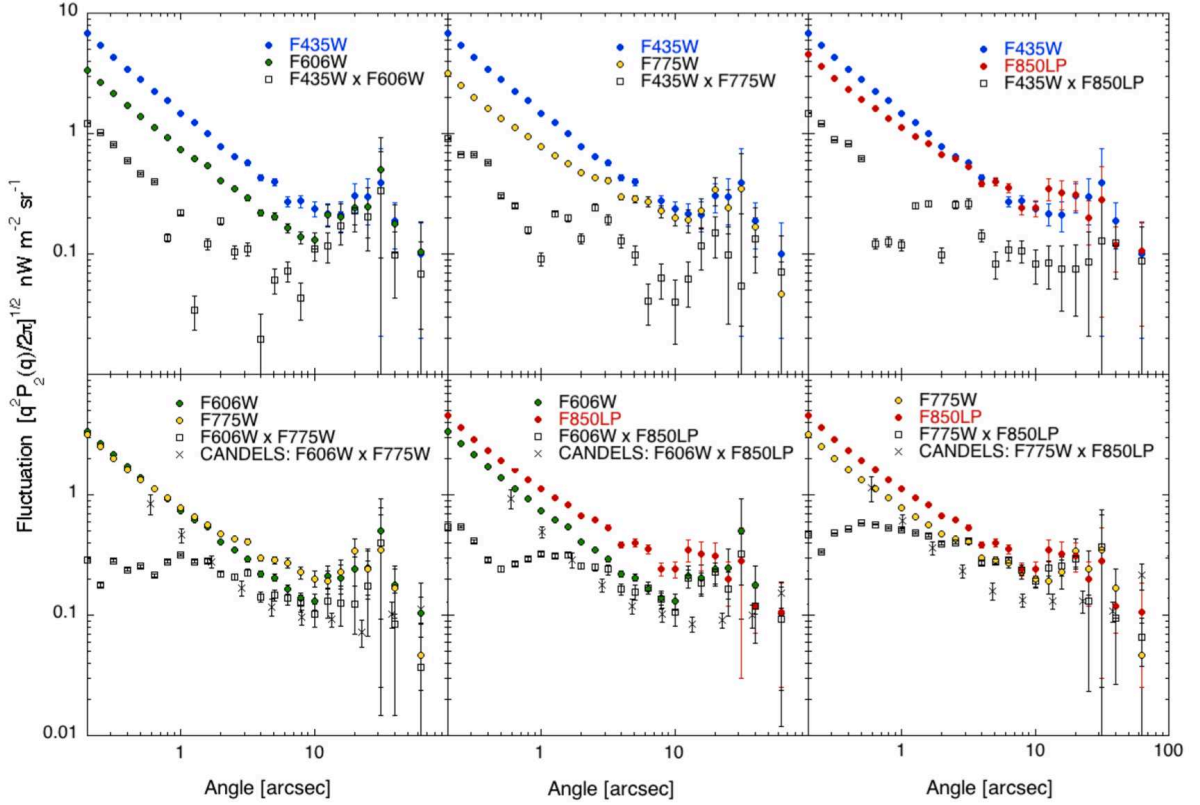} 
\end{center}
\caption{Cross-correlations for the XDF field. The squares represent the cross-correlation for the two wavelength bands
whose auto-correlations are given in Figure 3 in each panel.  The crosses represent cross-correlations for the CANDELS field
\citep{M-W15}.}
  \label{fig4}
\end{figure}

The correlations among the three longer wavelength bands are excellent. In particular, F775W $\times$ F850LP shows an almost perfect 
correlation at angles larger than 10 arcsec. The correlation spectra are marginally flat and extend down to 0.2 arcsec, 
indicating the existence of the sky structure even at the 0.2 arcsec scale. In  other words, the sky is filled with 
unknown objects, even at 0.2 arcsec scale. The surface number density for one object in 0.2 square arcsec corresponds
to 3.2 $\times$ 10$^{8}$ degree$^{-2}$. This value is a lower limit and is much larger than that of ordinary galaxies, 
as is shown in the next section. 

Large correlations imply that the considerable sky brightness is due to unknown objects. 
The auto-fluctuation spectra of the F775W and F850LP bands must be larger than the cross-correlation spectrum, F775W $\times$ F850LP, 
since there may be an additional uncorrelated fluctuation component in each band.
Therefore, we assume that the cross-correlation spectrum is the minimum of the auto-fluctuation spectra 
for both the F775W and F850LP bands.
We created simulated images with a fluctuation spectrum  equal to the cross-correlation spectrum 
(F775W $\times$ F850LP) and obtained dispersion of 1$\sigma$ for the brightness distribution. Using 10 trials, we obtained a 1$ \sigma$
value of $8.10 \pm 0.01$ nW m$^{-2}$ sr$^{-1}$. Since the sky brightness must always be positive, we take the 
3$\sigma$ brightness, 24 nW m$^{-2}$ sr$^{-1}$,
as the lower limit of the absolute sky brightness in the F775W and F850LP bands.
Moreover, we estimated the ILG fainter than the limiting magnitude based on the 
surface number density magnitude relation (black solid lines in Figure 7) and obtained  0.053 and 0.056  nW m$^{-2}$ sr$^{-1}$ for 
F775W and F850LP bands, respectively, which are negligible compared with the obtained lower limit. 
Furthermore, the detected lower limit is too bright to be explained by the IHL \citep{Cooray12b}.  

Figure 5 presents a summary of the observed excess brightness over the ILG. 
The obtained lower limit is indicated by a thick red horizontal bar, whose width covers the wavelength range of F775W and F850LP. 
The black solid line indicates the model ILG based on the deep survey and model galaxy \citep{Totani00, Matsumoto05}.
The results of other authors represent the excess brightness after subtracting the model ILG from the observed EBL,
depending on the limiting magnitudes. 
Figure 5 indicates that our findings are
consistent with results acquired for  CIBER/LRS \citep{Matsuura17} and HST \citep{Bernstein07}.
\citet{Zemcov17} analyzed data from the Long Range Reconnaissance Imager instrument on NASA's New Horizons
mission for the wavelength range of 440 - 870 nm and reported an upper limit, 19.3 nW m$^{-2}$ sr$^{-1}$, beyond the Jovian orbit, which is 
lower than our lower limit. However, their result does not contradict our findings, since there exists a large systematic error, 
+10.3 nW m$^{-2}$ sr$^{-1}$, in their result.  
\citet{Mattila17} performed a spectroscopy study of the stellar Fraunhofer lines toward a dark cloud at high galactic latitudes, and 
estimated the EBL by comparing the line strength measured in and outside the dark cloud. The authors
reported the EBL detection at 400 nm with an upper limit at 520 nm (green symbols in Fig. 5), and
reported that the detected EBL was
too high to be explained by the ILG. The excess brightness detected in that study is slightly lower
than the value in this work, however, two results are not inconsistent, considering the large errors and the difference in wavelengths.
\citet{Kawara17} reported the detection of isotropic emission at 0.2 - 0.6 $\mu$m using the HST/FOS sky spectra, which
is consistent with the EBL detected by \citet{Bernstein07}. The authors attributed the isotropic emission to
the solar system origin, because the spectrum is similar to that of the zodiacal light. However, it is possible  that
part of the detected isotropic emission is EBL. 
\citet{Matsuoka11} analyzed the data from Imaging Photopolarimeter onboard Pioneer 10 and obtained the optical EBL beyond the
Jovian orbit, which is
consistent with the ILG. However, we did not include their data in Figure 5 because they did not account for the effect of instrumental 
offset \citep{Matsumoto18}.

Thus, in summary, our analysis reveals that there exists bright excess background light, even at optical wavelengths, and
that the energy density in the optical and near-infrared wavelengths is a few times or more higher than
that previously thought (black line).

\begin{figure}
 \begin{center}
  \includegraphics[width=16cm]{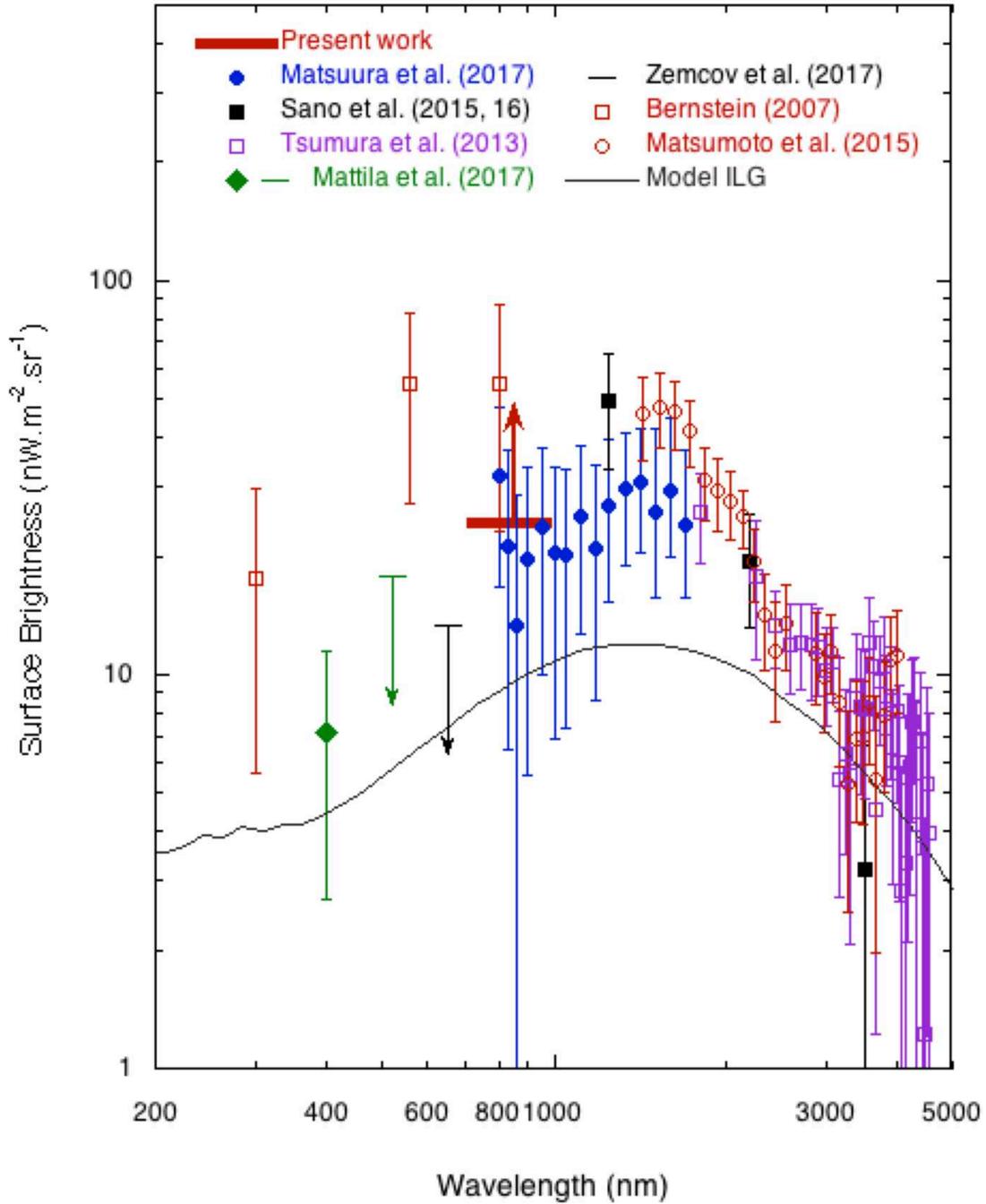} 
\end{center}
\caption{Summary of observations of the optical and near-infrared excess background over the integrated light of galaxies (ILG). 
The thick red horizontal bar indicates the lower limit we obtained. The black solid line represents the model ILG which includes 
all galaxies \citep{Totani00, Matsumoto05}}
  \label{fig5}
\end{figure}

\section{Fluctuation source}
\subsection{Faint compact objects (FCOs)}
Fluctuation analysis reveals the presence of unknown celestial objects that are densely distributed throughout the sky.
To identify these unknown objects, we carefully examined XDF images and found that
the density of compact objects dramatically increases toward the faint end. We examined the nature of these objects
based on the catalog of UDF objects reported by \citet{Beckwith06}, since a XDF catalog is not available.

In Figure 6, we present UDF objects in a magnitude/stellarity diagram, where objects that were identified as stars in \citet{Pirzkal05} are removed. 
In Figure 6, 8918 objects are plotted for the F775W band for which stellaities are 
measured in the UDF catalog. 
The stellarity index in SExtractor was developed to identify stars in images  \citep{Bertin96} by
evaluating the compactness and the presence of an extended envelope. Objects with a stellarity value larger than 0.8 are usually
regarded as stars. Figure 6 shows a group of high-stellarity objects in the faint region which we denote as 
FCOs. In Figure 6, the fainter-magnitude FCOs have lower stellarities, however, this trend is most likely 
not an intrinsic feature. Confusion with the sky fluctuations extending down to 0.2 arcsec scale is more problematic
for faint objects, which reduces their apparent stellarity.

\begin{figure}
 \begin{center}
  \includegraphics[width=12cm]{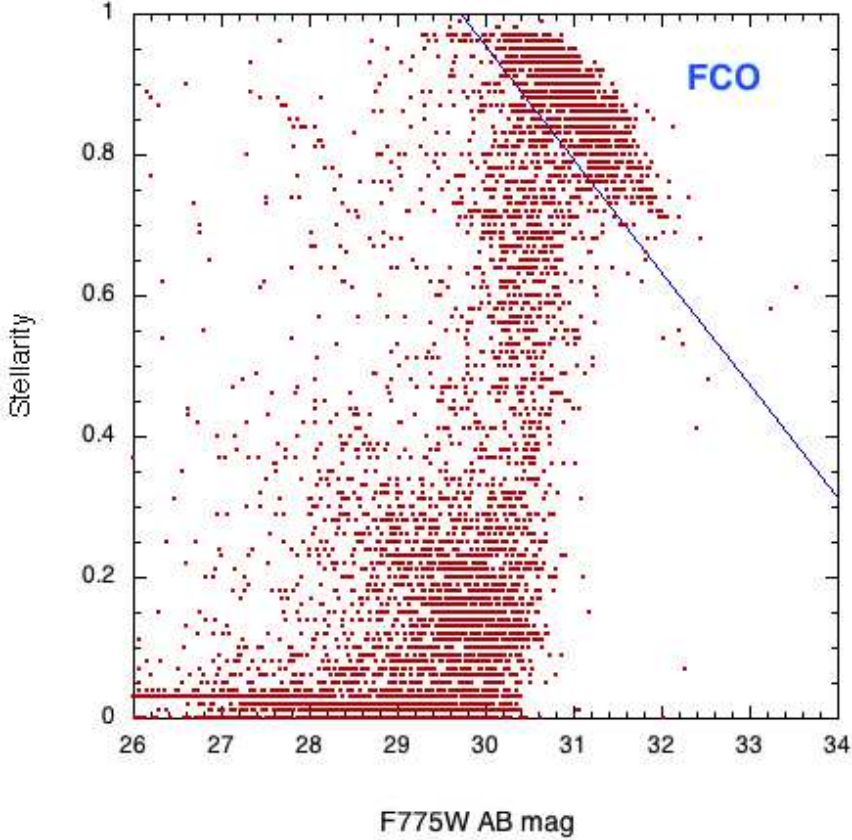} 
\end{center}
\caption{Stellarity/magnitude (F775W) diagram of 8918 objects. Data points for objects with the same values are overlaid. 
The blue solid line is 
drawn rather arbitrarily to separate FCOs from ordinary galaxies.}
   \label{fig6}
\end{figure}

A somewhat arbitrary line is drawn in Figure 6 to separate FCOs from ordinary
galaxies. The number of FCOs, defined as objects located to the upper right of the solid line, is 1283.
The catalog of the detected FCOs is given in Supplementary Data on line.
For a detailed analysis, we retrieved 649 FCOs whose stellarities are larger than 0.9.

Figure 7 shows the cumulative number counts as a function of the apparent magnitude, $m$. The filled squares and filled circles 
represent all objects and objects with stellarity $> 0.9$, respectively.  Figure 7
clearly shows that the surface number density of FCOs increases 
much more dramatically than that of all objects toward the faint end. We fit the slopes for FCOs in the 
F775W band using a power law (red lines) such that the
residual numbers of objects in the bright region are constant (open squares). The result, $\propto$ 10$^{1.2m}$, is much steeper than the isotropic
distribution, $\propto$ 10$^{0.4m}$, and corresponds to the density distribution, $n(r)\propto r^{3}$, where $n(r)$ is the spatial
number density and $r$ is the radial distance.
The surface number density of FCOs  must peak at a certain magnitude; otherwise the sky brightness would diverge.  

The F606W and F850LP bands show a sharp increase, similar to that of the F775W band, but
the slopes are lower than that of F775W, most likely because of color scattering.
For the F435W band, it is difficult to separate the FCOs because ordinary galaxies with large 
stellarities  are not negligible, even at 30 mag.
This trend may account for why shot noise remains for the cross-correlation with the F435W band in Figure 4. 
We extrapolated the slopes of all objects toward the faint region to be $\sim$ 10$^{0.244m}$ by applying the
evolution model \citep{Helgason12}, which was used to estimate the fluctuation of ordinary galaxies in Figure 2. 
Because the model predicts smooth slopes up to 32 mag, the
turn over of all the detected objects are due to the detection limit. The turn overs of the FCOs
occur at slightly fainter magnitudes than those for all objects because the detection limit of a
compact object is better than that of an extended object. If we assume a simple extrapolation
for FCOs and consider the total number of FCOs to be twice the value given in Figure 7, the surface number densities of
FCOs exceed those of ordinary galaxies at 31.8 mag for the F775W band.
If we assume that the excess surface brightness is due to the integrated light of FCOs, the lower limit of the 
surface brightness of 24 nW m$^{-2}$ sr$^{-1}$
implies a cut-off magnitude of 34.9 for the F775W band. In this case, the cumulative surface number density
is determined as $3.35 \times 10^{10}$ (degree)$^{-2}$, $2.6 \times 10^{3}$ (arcsec)$^{-2}$, or $1.38 \times 10^{15}$ for the entire sky.

\begin{figure}
 \begin{center}
  \includegraphics[width=16cm]{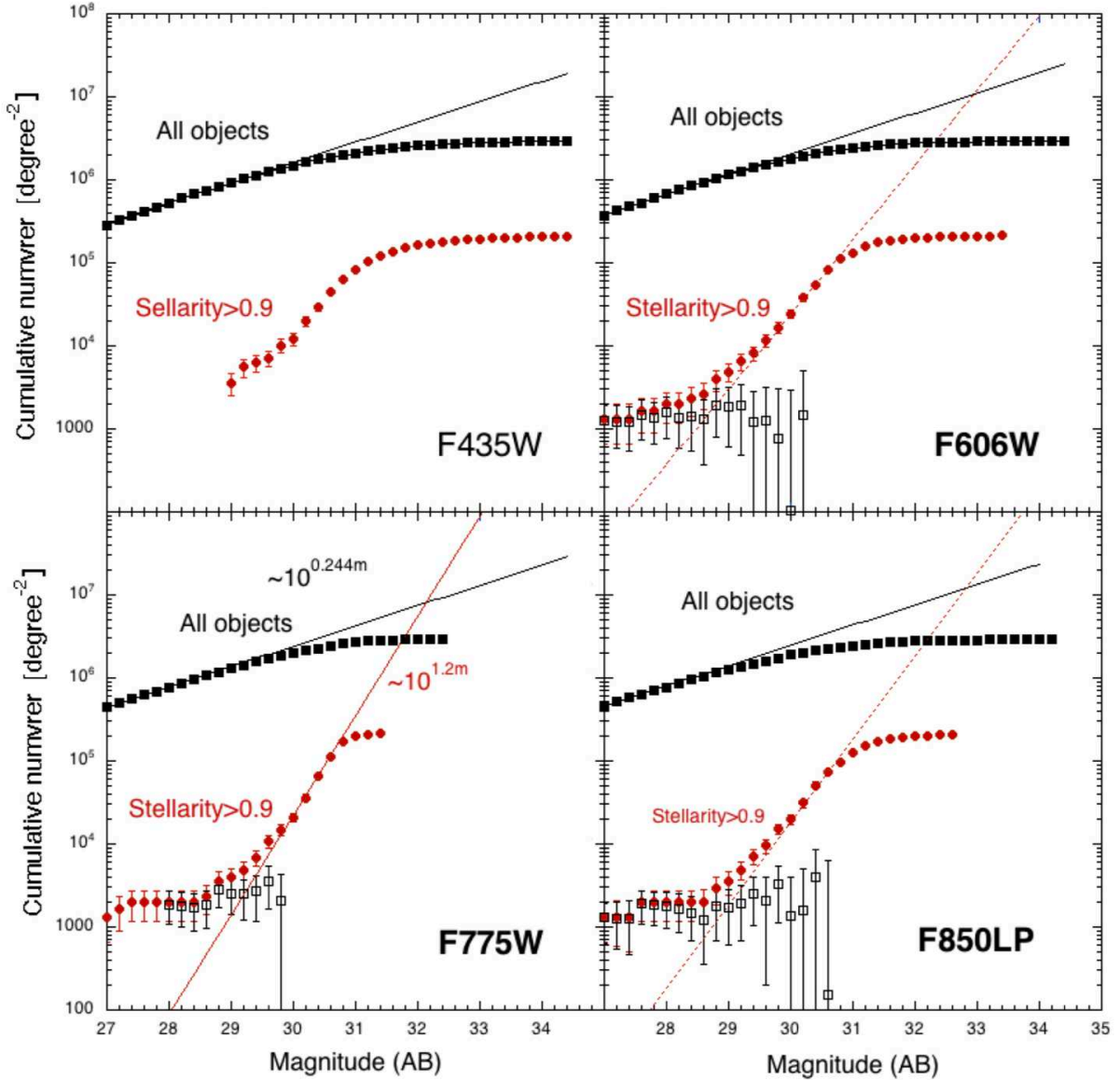} 
\end{center}
\caption{The cumulative surface number densities of four optical bands as a function of the apparent magnitude, $m$. 
The filled squares and filled circles represent all objects and objects with stellarity $> 0.9$, respectively. The open squares denote the
surface number density of ordinary galaxies with stellarity $> 0.9$.}
   \label{fig7}
\end{figure}

To evaluate the validity of using stellarity index, we examined the compactness of FCOs using images with a pixel resolution of 30 mas.  
We divided the images into annular rings with a width of
0.03 arcsec and obtained the average
surface brightness in units of $\lambda \cdot I_{\lambda}$: nW m$^{-2}$ sr$^{-1}$ for each ring. 
Using the same procedure, we obtained radial profiles of the PSF using stellar images for comparison with those for the FCOs.
Figure 8 shows a typical example of the radial profiles and XDF images for a bright FCO, UDF8942, which
are not distinguishable from stellar images for all wavelength bands.

\begin{figure}
 \begin{center}
  \includegraphics[width=14cm]{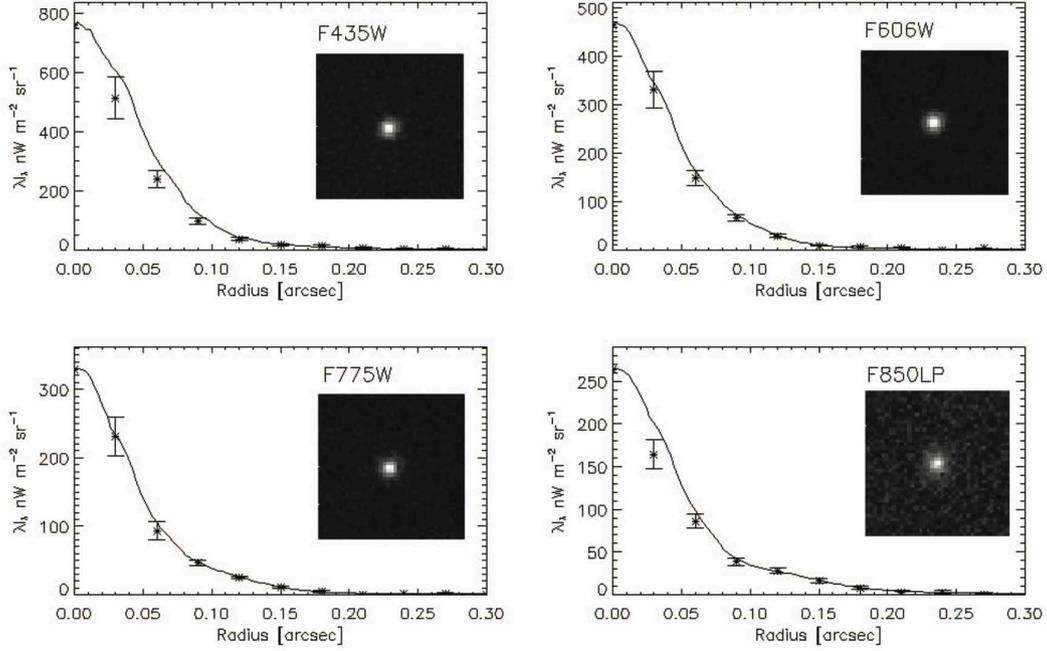} 
\end{center}
\caption{Radial profiles and XDF images of a bright FCO, UDF8942, for four optical bands. The solid lines represent the PSF
obtained from stellar images.  Image scales are $41 \times 41$ $pixel^{2}$, or $1.23 \times 1.23$ $arcsec^{2}$.}
   \label{fig8}
\end{figure}

Furthermore, we selected 48 bright FCOs as an order of the average errors for the four wavelength bands and examined their radial profiles
in the combined F606W and F775W images, since the signal-to-noise ratios for the F435W and F850LP bands are relatively poor.
The obtained radial profiles for the 48 bright FCOs are shown in Figure 20 in Appendix 2. 
These profiles are highly similar to the PSF, indicating that the stellarity in the UDF catalog is a reliable index for identifying the compact sources.
The two bright objects, \#1 (UDF9397) and \#3 (UDF6732), are identified as quasars (Table 1) whose radial profiles are fairly consistent with
the PSF, which confirms the validity of our analysis.  
Objects \#2 (UDF7025), \#4 (UDF4070), \#5 (UDF8230),  \#6 (UDF5711), \#8 (UDF2087) and \#12 (UDF5550) show a slight enhancement 
at 0.1 - 0.2 arcsec, which suggests that they may be ordinary galaxies with an extended halo or disk. 
 Object \#36 (UDF1240) shows a deviation of the radial profile from the 
PSF, which could be ascribed to source confusion, as there exists another source in the immediate vicinity. 
For the other FCOs, the radial profiles (Figs. 8, 20) are not distinguishable from the PSF within the errors, and 
extended envelopes are not detected.

The detection limits of the envelope surface brightness are determined for the structure of the sky, whose
3$\sigma$ fluctuation level for $0.12 \times 0.12$ arcsec$^{-2}$ amounts to 
$\lambda \cdot I_{\lambda}$ $\sim$ 5.3, 2.58, 2.52 and 3.58  nW m$^{-2}$ sr$^{-1}$
for the F435W, F606W, F775W and F850LP band, respectively,
which correspond to 28.3, 28.7, 28.5 and 27.9 mag(arcsec)$^{-2}$. 
The exponential disk of spiral galaxies has a much brighter surface brightness ($<26$ mag(arcsec)$^{-2}$) for scales larger than 1 kpc. 
Since the minimum angular scale of 1 kpc corresponds to 0.12 arcsec at $\it z \sim$1.6, the radial profiles of FCOs 
indicate that extended disks do not exist in the FCOs except for the six bright cases.
However, we must also consider that the surface brightness rapidly decreases at high redshifts.

To qualitatively estimate the compactness of FCOs, we calculated convolved radial profilers with the PSF for
dwarf spheroidals, which follow de Vaucouleur's law,
$$
I(r)=I_{e} \exp(-7.67((r/r_{e})^{1/4}-1)) 
$$
where $r_{e}$ is the effective radius in which half of the total light is contained. 
Figure 9 shows a comparison of the PSF for the F775W band and the convolved images
for $r_{e}$=0.02, 0.05. 0.1 and 0.2 arcsec.
Figure 9 indicates that a larger $r_{e}$ correspond to a more extended envelope, while the $r_{e}$ of the FCOs is at most 0.02 arcsec.
Since the angular distance has a maximum at $\it z \sim$1.6, where 0.1 arcsec corresponds to 860 pc, the physical size
of $r_{e}$ must be smaller than 172 pc. This size corresponds to the low end of the size range of ordinary dwarf spheroids  \citep{Zhang17}.

\begin{figure}
 \begin{center}
  \includegraphics[width=12cm]{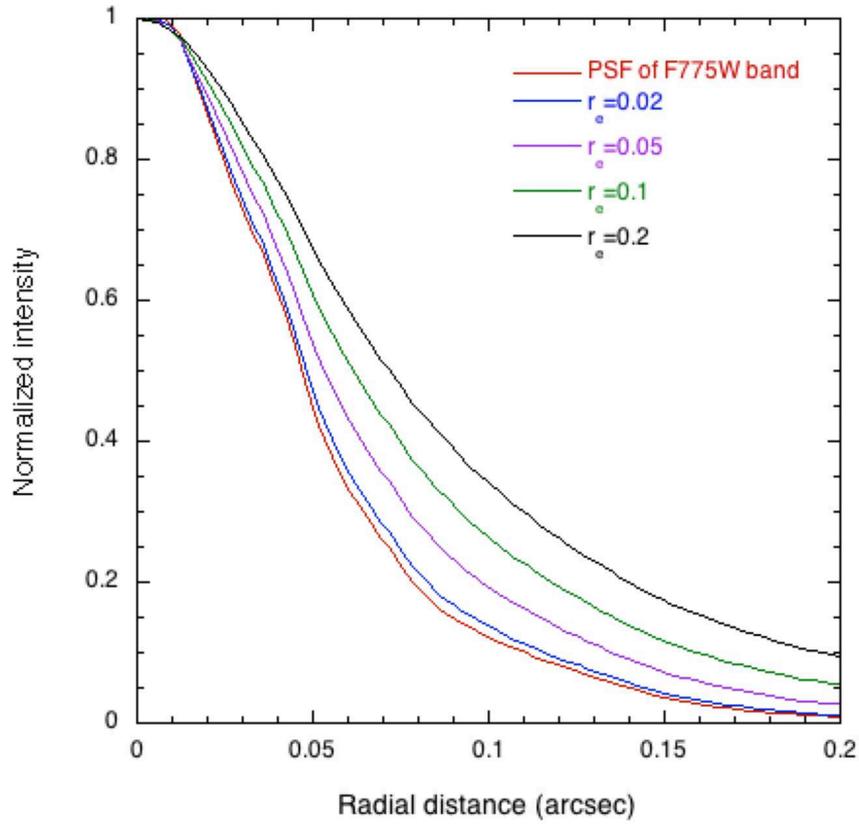} 
\end{center}
\caption{PSF(red lines) for the F775W band compared with convolved images of the radial distribution of dwarf spheroidals 
that follow de Vaucouleur's law for different effective radii, $r_{e}$.}
   \label{fig9}
\end{figure}

As a next step, we examined the photometric data.   
Figure 10 presents a color/color diagram of 128 FCOs whose average errors are less than 0.2 mag.
The left and right panels show (F606W-F775W)/(F435W-F606W) and (F775W-F850LP)/(F606W-F775W), respectively. 
The straight lines indicate the loci of a blackbody whose temperature ranges from 4000 K to 14000 K. 
Although a clear sequence is not found due to the large errors, Figure 10
indicates that the color temperatures are fairly high ($\sim$ 10000 K), and the centroids are close to the point of origin.
This result implies that the magnitudes of the four optical bands are similar, that is, 
the SED in units of $\lambda \cdot I_{\lambda}$ shows a dependence on $\lambda^{-1}$.

\begin{figure}
 \begin{center}
  \includegraphics[width=16cm]{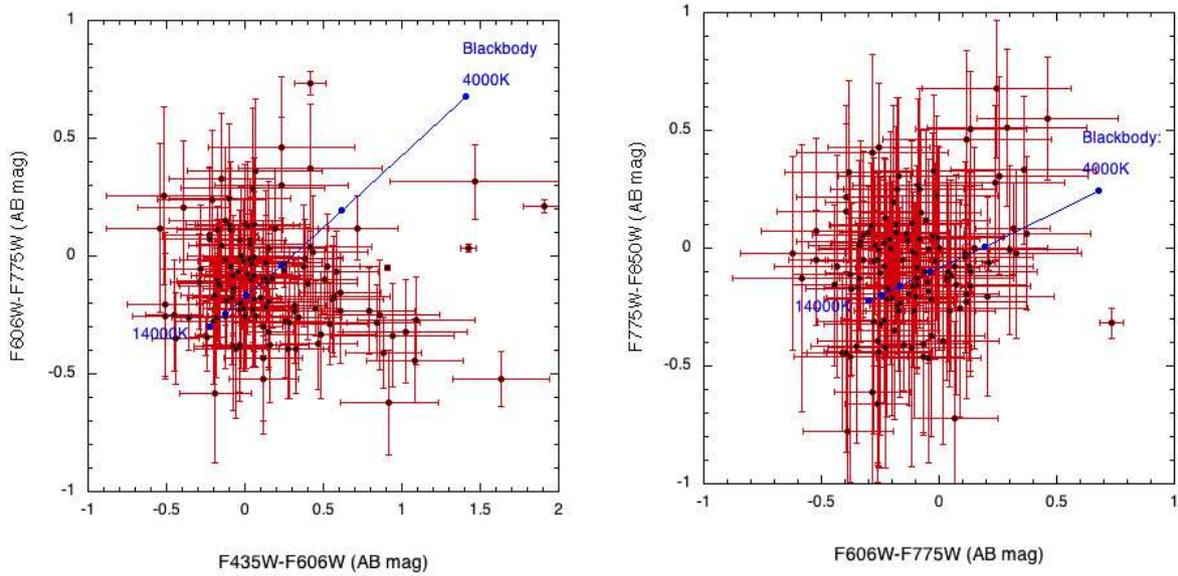} 
\end{center}
\caption{ Color-color diagrams for bright FCOs. The blue solid lines represent the loci of the blackbody.}
   \label{fig10}
\end{figure}

The SED is a key factor in understanding the nature of FCOs. Since the fluxes of only four optical bands are presented
in the UDF catalog, we additionally used the UVUDF catalog \citep{Rafelski15}, which provides fluxes for three UV and
four infrared bands, in addition to the optical bands. Furthermore, we independently attained photometry for the Spitzer images, 
however, the number of detected objects is limited by source confusion due to the large beam size of Spitzer.
Figure 11 presents the typical SEDs of four FCOs. The SEDs for the 48 bright FCOs used in analyzing the radial profiles 
are also shown in Figure 21 in Appendix 2.

The SEDs in the optical bands are featureless, exhibiting a power law behavior as 
expected from the color-color diagram (Fig. 10). The observed optical SEDs are typically found for starburst galaxies and AGNs. 
In contrast, excess structures over the power law are found
in the infrared bands. Interestingly, the SEDs exhibit a drastic change at the borderline at 1 $\mu$m for almost all bright
FCOs.

\begin{figure}
 \begin{center}
  \includegraphics[width=14cm]{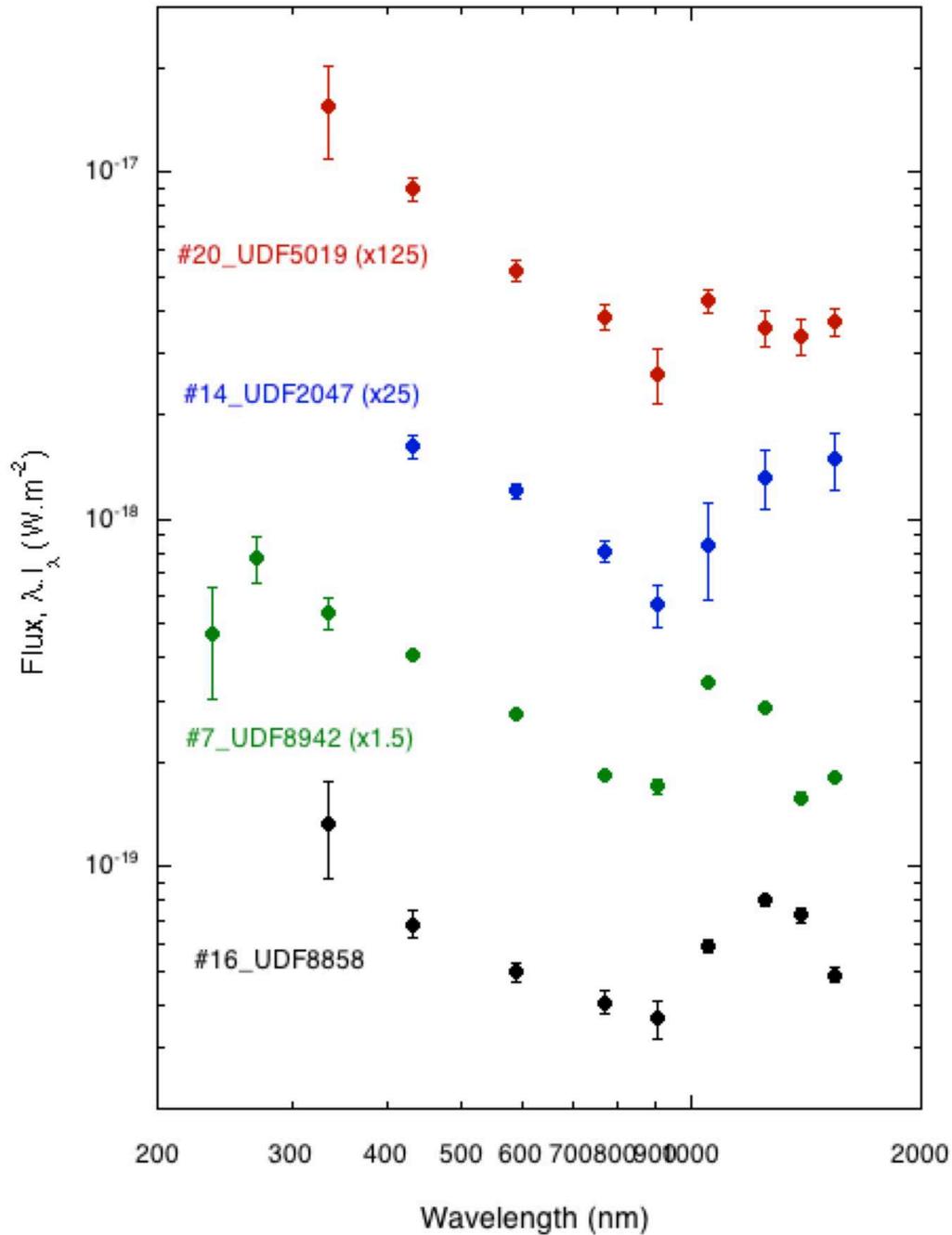} 
\end{center}
\caption{SEDs of typical FCOs based on the UVUDF catalog \citep{Rafelski15}. For clarity, the flux scales are magnified as shown in parenthesis.}
   \label{fig11}
\end{figure}

\citet{Rafelski15} estimated the photometric
redshifts of UDF objects using two evolution models of the galaxy, BPZ and EAZY. 
The results for 
UDF objects for which spectroscopic observations have been made and those in Figure 11 are shown in Table 1, as well as
references for previous observations. 

\begin{landscape}
\begin{table}[h]
  \caption{Information from bibliographies. } 
  \label{bibliographies}
  \begin{tabular}{ccccccccccc} \hline
  $ ID \sharp$ & UDF $\sharp$  & F775W mag$^a$ & z BPZ$^a$ & z EAZY$^a$ & Emission line detected by MUSE$^b$ & z MUSE$^b$ & CONF.$^b$& Identification \\ 
  \hline
  1  & 9397  & 21.031 $\pm$ 0.001   & {\scriptsize 1.30} -1.42- {\scriptsize 1.52} & {\scriptsize0.12}-0.17-{\scriptsize1.33} & H$\delta$,OII1                                                 & 1.2185 & 2 & quasar$^c$ \\
  2  &	7025   & 25.276 $\pm$ 	0.006    & {\scriptsize3.21}-3.40-{\scriptsize3.59}  & {\scriptsize2.89}-2.96-{\scriptsize3.04} & Ly$\alpha$                                                        & 3.326   & 2 & galaxy$^d$ \\
  3  &	6732   & 24.587 $\pm$  0.006    & {\scriptsize2.90}-3.09-{\scriptsize3.27}  & {\scriptsize3.13}-3.19-{\scriptsize3.25} & CIII1,Ly$\alpha$                                              &3.1882  & 2 & quasar$^e$ \\
  4  & 4070  & 25.959 $\pm$  0.01     & {\scriptsize2.44}-2.60-{\scriptsize2.75}  & {\scriptsize1.76}-1.84-{\scriptsize1.91} &CIII1,CIII2                                                         & 2.4908 & 3 & galaxy$^{d, f}$ \\
  5  & 8230  & 25.531 $\pm$  0.013   & {\scriptsize3.34}-3.54-{\scriptsize3.73}  & {\scriptsize3.14}-3.22-{\scriptsize3.29} & Ly$\alpha$                                                        & 3.6716 & 3 & galaxy$^{d, f}$ \\
  6  & 5711  & 27.353 $\pm$  0.016   & {\scriptsize0.51}-0.59-{\scriptsize0.77}  & {\scriptsize0.09}-0.12-{\scriptsize0.17} & H$\beta$,H$\gamma$,OII1,OII2, OIII1,OIII 2	& 0.6206 & 3 & galaxy$^{g, h}$ \\  
  7  & 8942  & 27.662 $\pm$  0.033   & {\scriptsize0.70}-0.78-{\scriptsize0.85}  & {\scriptsize0.01}-0.03-{\scriptsize0.05} & OIII2                                                                  & 1.4469 & 2 & \\	
  8  & 2087  & 26.597 $\pm$  0.014   & {\scriptsize3.50}-3.72-{\scriptsize3.94}  & {\scriptsize3.33}-3.42-{\scriptsize3.51} & Ly$\alpha$                                                        & 3.7282 & 2 & galaxy$^d$ \\
 12 & 5550  & 28.781 $\pm$  0.0533 & {\scriptsize2.60}-2.78-{\scriptsize2.95}  & {\scriptsize2.48}-2.61-{\scriptsize2.80} &                                                                          &              &    & \\				
 14 & 2047  & 29.103 $\pm$  0.0708 & {\scriptsize1.50}-1.96-{\scriptsize2.35}	& {\scriptsize1.66}-2.04-{\scriptsize2.42} &                                                                          &              &    & \\					
 16 & 8858  & 28.856 $\pm$  0.078   & {\scriptsize1.54}-1.69-{\scriptsize1.84}  & {\scriptsize1.59}-1.70-{\scriptsize1.80} &                                                                          &              &    & \\					
 20 & 5019  & 29.155 $\pm$  0.0886 & {\scriptsize1.27}-1.52-{\scriptsize2.12}  & {\scriptsize1.37}-1.57-{\scriptsize1.89} &                                                                          &              &    & \\
 \hline                                                 
  \end{tabular}
  \begin{tabnote}
  1. CONF. is a confidence level of the MUSE detection in which 1 corresponds to the best case. \\
  2. References are follows. 
  a: \citet{Rafelski15}, b: \citet{Inami17}, c: \citet{Veron10}, d: \citet{Straatman16}, e: \citet{Fiore12}, \\ f: \citet{Camero11}, g: \citet{Thompson05},h: \citet{Coe06}
  \end{tabnote}
\end{table}
\end{landscape}

\subsection{What are FCOs?}
We first, examined the known objects similar to FCOs. Globular clusters associated with masked galaxies may be
a candidate of FCOs. However, globular clusters have a difficulty to reproduce the steep slope for the dependence 
of the surface number density on the magnitude in Figure 7, since globular clusters should show a similar slope to the ordinary galaxies. 
Furthermore, SEDs of FCOs are quite different from those of globular clusters which consist of
late type stars. These evidences indicate that globular clusters cannot be FCOs. 

With respect to compactness, ultra-compact dwarfs (UCDs) are similar to FCOs. However, their SEDs 
resemble those of late type galaxies (\citet{Drinkwater00}, \citet{Zhang17}), 
which differ from those of FCOs. UCDs are thought to be formed in the the cluster of galaxies 
by stripping the outer envelopes via collisions with galaxies. UCDs are few in number and occur in local; thus,
they cannot be FCOs.

The third candidate is a dwarf galaxies undergoing an intense starburst, known as extreme emission line galaxies (EELGs). 
Such galaxies at a redshift of $\sim$2 are characterized by strong [OIII] and [NII] line emissions, 
which show excess infrared emission  (\citet{Wel11}, \citet{Maseda13}, \citet{Maseda14}). 
FCOs exhibit a similar SED, however, the luminosity of FCOs is much lower than that of EELGs, if FCOS are EELGs.

\citet{Inami17} attained spectroscopic data for UDF objects using the Multi-Unit Spectroscopic Explorer (MUSE) on VLA and
detected emission lines for eight UDF objects (Table 1). The redshifts obtained for seven of these objects which have 
already been cataloged as 
quasars or galaxies, are reasonably consistent with the photometric redshifts obtained by \citet{Rafelski15}.
For the remaining object, \#7 (UDF8942),
the MUSE result presented  [OII] line emission, and the redshift of 1.447 was assigned, however, 
\citet{Rafelski15} obtained photometric redshifts of
0.78 (BPZ) and 0.03 (EAZY). The line intensity of [OII] detected by MUSE is 
6.74 $\pm$ 1.9 $\times$ 10$^{-22}$ Wm$^{-2}$, which is close to the detection limit, and the redshift was determined from only this line.
Figure 12 shows the SED of \#7 (UDF8942) and the positions of emission lines expected for a redshift of 1.447. 
 We estimated the line intensity of [OIII] by assuming [O III]/[O II] $\sim$10, which
 is the maximum case for the observed Lyman $\alpha$ emitters and Lyman break galaxies \citep{Nakajima16}. 
 Even in this case, the contribution of the [OIII] line to the F125W band is 5.6 $\times$ 10$^{-20}$ Wm$^{-2}$, which is much 
 lower than the excess flux. 
No corresponding emission line exists for the excess in the F106W band, and no indication of H$\alpha$ emission is observed
for the F160W band. Furthermore, the upper limits of the Spitzer observation exclude the possibility of a Balmer jump for the 1 $\mu$m excess. 
Based on these results, we conclude that the redshift of \#7 (UDF8942) is not yet determined and that the origin of excess in
the F105W and F125W bands is still uncertain. Thus, it is difficult to explain the 
SED of  \#7 (UDF8942), as well as SEDs of other bright FCOs (Figs. 11 and 21), using emission lines.

\begin{figure}
 \begin{center}
  \includegraphics[width=12cm]{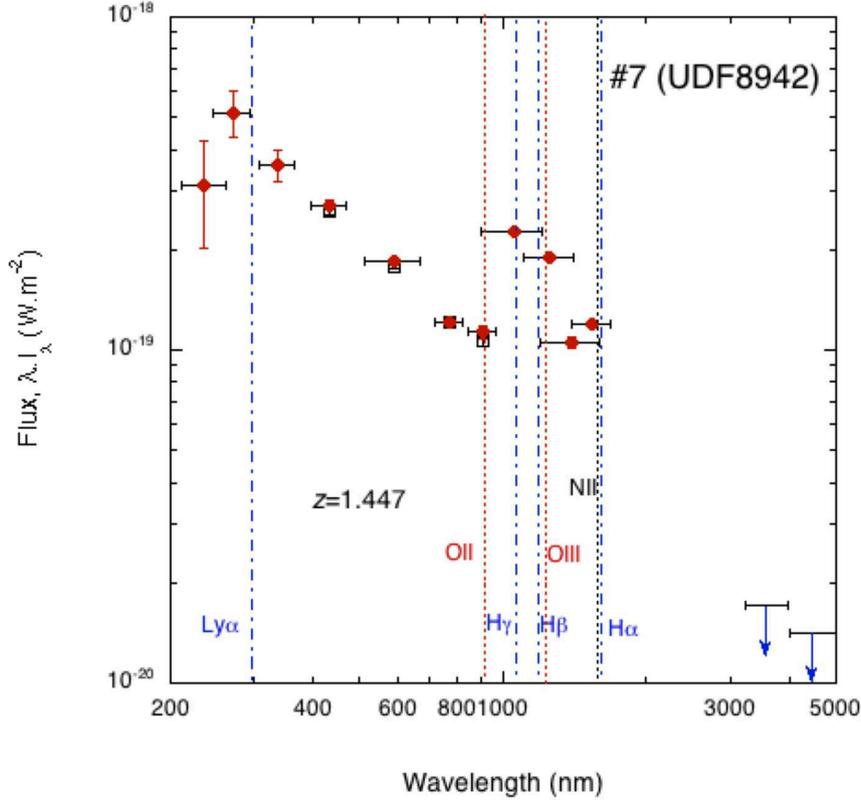} 
\end{center}
\caption{SED of \#7 (UDF8942) compared with the positions of emission lines (filled diamonds and dotted lines), assuming a redshift of 1.447,
as reported for MUSE. The red filled circles, open squares, and blue arrows represent the fluxes of the UVUDF, and UDF catalogs 
and the upper limits measured in this work for 
Spitzer, respectively. The horizontal bars are not error bars; rather indicate the bandwidths of the filters.}
\label{fig12}
\end{figure}

Since there exists no clear corresponding object for FCOs, we examined the characteristic features of FCOs more carefully.
Figure 13 shows the correlation between photometric redshifts obtained from BPZ and those from EAZY  \citep{Rafelski15}, 
in which two quasars are not included.
The errors for EAZY are smaller than those for BPZ, however, the uncertainties are fairly large. The redshift ranges from 1.5 to 3.5 in
both cases. In the reported analysis, a lower redshift is defined so that the Lyman $\alpha$ line (121.5 nm) remains in the F435W band, while
a larger redshift is obtained by attributing a Balmer jump (364.6 nm) to infrared excess.  However, the SEDs in Figures 11 and 21 show 
different features from that expected for the Balmer jump. Thus, we suggest that the photometric redshifts of FCOs obtained by \citet{Rafelski15} 
may not be accurate.

\begin{figure}
 \begin{center}
  \includegraphics[width=12cm]{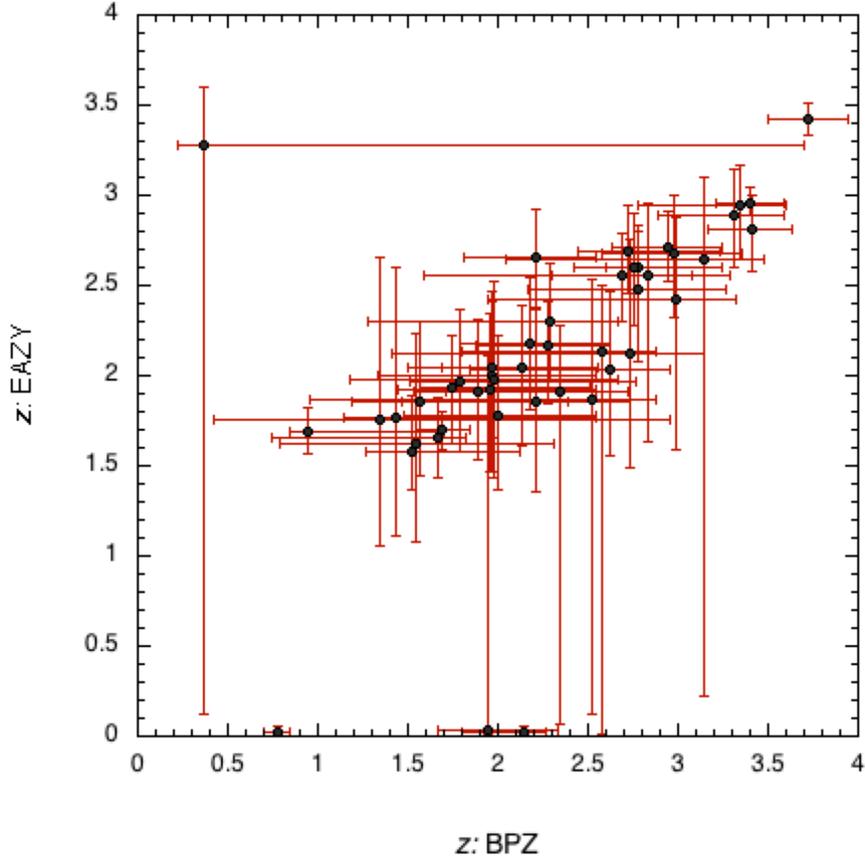} 
\end{center}
\caption{Correlation diagram for the redshifts obtained from BPZ and EAZY. The error bars represent the allowable range
of redshifts listed in the UVUDF catalog \citep{Rafelski15}.}
\label{fig13}
\end{figure}

Figure 14 presents the magnitude/redshift relation for objects from the UVUDF catalog \citep{Rafelski15}. 
The left and right panels show data for FCOs (stellarity $> 0.9$) and randomly chosen ordinary galaxies, respectively. 
An anomalous features is observed for the FCOs that is not 
expected for randomly distributed objects. A simple explanation for Figure 14 is that the redshifts of the FCOs (stellarity $> 0.9$)
are almost the same; thus, FCOs (stellarity $> 0.9$) are localized at a fixed redshift. This trend is consistent with the 
SEDs of FCOs (Figs. 11 and 21), 
where an SED change at 1 $\mu$m
is commonly found. From Figure 21, the redshift range of the FCOs is estimated to be $\Delta z/z < 0.1$.

\begin{figure}
 \begin{center}
  \includegraphics[width=16cm]{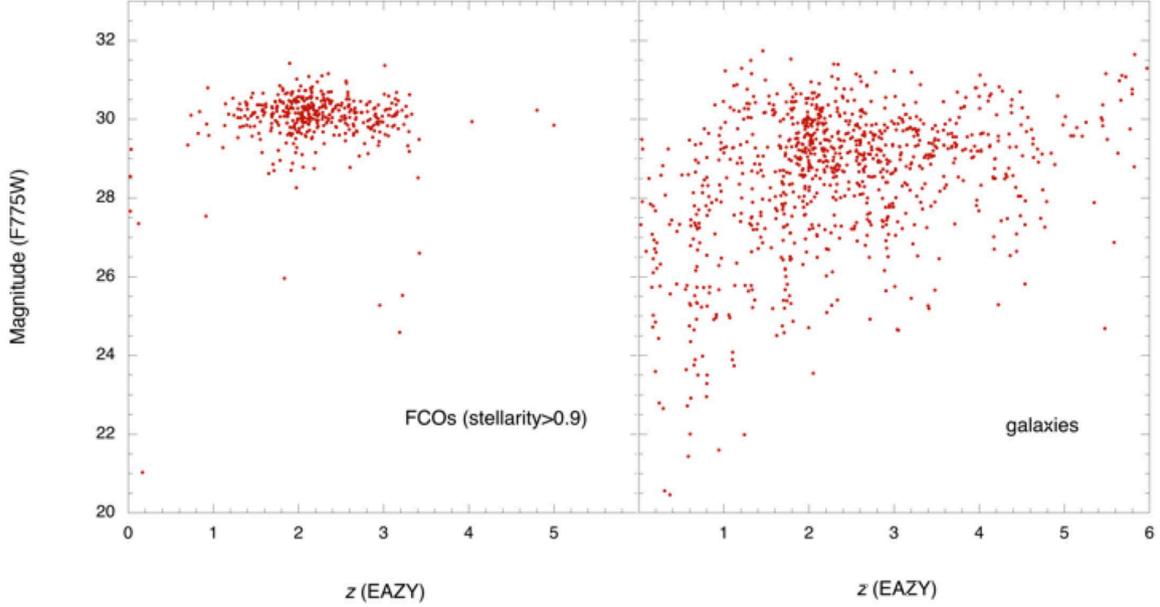} 
\end{center}
\caption{The redshift (EAZY) dependence on the magnitude at F775W. The left panel shows that for FCOs of stellarity
larger than 0.9, while the right panel indicates that for the ordinary galaxies randomly chosen from the UVUDF catalog \citep{Rafelski15}.}
\label{fig14}
\end{figure}

To delineate the nature of FCOs, we used two simplified models. We first assumed that FCOs are located at
a fixed redshift, $z_{c}$ ($\Delta z/z_{c} < 0.1$), and have 
a luminosity function of $\propto 10^{1.2m}$; same shape as surface number density magnitude relation in Figure 7 (model 1). 
In this model, 
FCOs appear at certain epochs and disappear after a certain lifetime. The luminosity function can be attributed to
the mass difference or elapsed times. 
We estimated the total flux due to FCOs based on the lower limit of the surface brightness for the F775W band, 24 nW m$^{-2}$ sr$^{-1}$.
Assuming that the optical SED, $\lambda \cdot I_{\lambda}$, $\propto \lambda^{-1}$ (Fig. 10) and that contribution of the sum of the
UV and infrared bands is the same as that for the optical band, we obtained an integrated surface brightness to be 47 nW m$^{-2}$ sr$^{-1}$. 
Since the number of FCOs for the entire sky is $1.38 \times 10^{15}$, we can estimate a lower limit of 
the average FCO luminosity that is dependent on $z_{c}$.

Furthermore, we assumed that FCOs consist of missing baryons, corresponding to half of the baryon 
mass ($2.8 \times 10^{9}$ M$_{\odot}$ (Mpc)$^{-3}$).
We can then obtain the average mass of FCOs and the mass-to-luminosity ratio, depending on $z_{c}$. 
These result are shown in Figure 15. 
The mass and luminosity of FCOs are significantly lower than those of dwarf galaxies, while the mass-to-luminosity ratio
of FCOs at $z_{c}  \sim 1$ is a permissible range as dwarf galaxies.
However, mass-to-luminosity ratio
in Figure 15 is over-estimated because the total luminosity is at the lower limit and the existence of faded FCOs is not taken into account.

\begin{figure}
 \begin{center}
  \includegraphics[width=12cm]{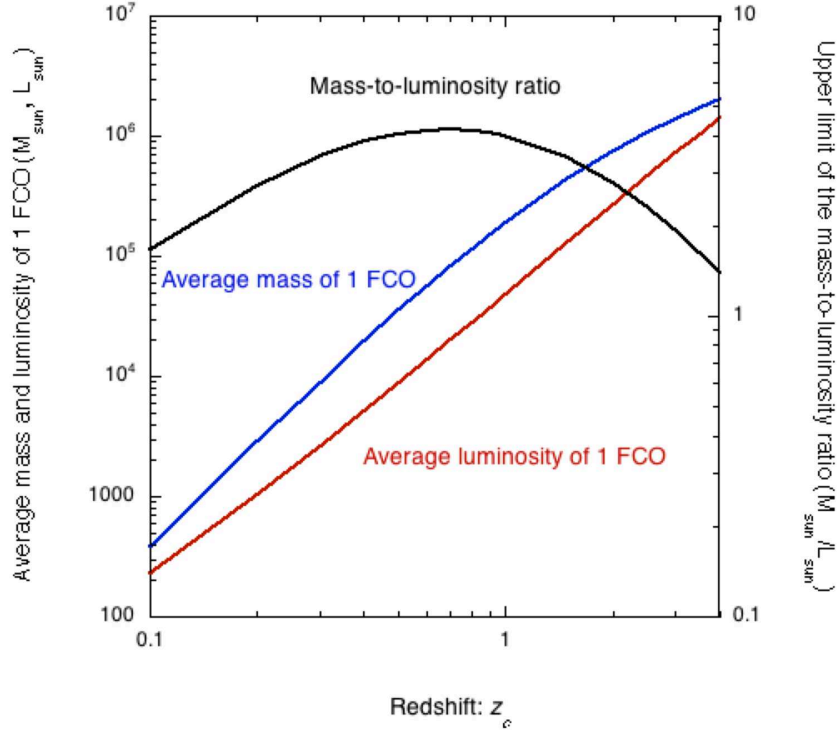} 
\end{center}
\caption{ Dependence of the average luminosity and mass (left ordinate) and mass-to-luminosity ratio (right ordinate) on 
the redshift, $z_{c}$, based on model1. }
\label{fig15}
\end{figure}

For the second simplified model, we assumed that the luminosity of FCOs is intrinsically constant (model 2). 
In this case we attributed the 
dependence of the surface number density on the apparent magnitude (Fig. 7) to the radial distribution of FCOs. 
We applied the spatial density distribution, $n(r)\propto r^{3}$, where $n(r)$ is the spatial number density of 
FCOs and $r$ is the comoving radial distance,
and assumed that FCOs do not exist beyond $r_{max}$ ($z_{max}$). 
This model is the simplest case that satisfies the dependence of the surface number density on the apparent magnitude in Figure 7.
We attributed the mass of FCOs to missing baryons and applied 
the same procedure to obtain the integrated luminosity employed in model 1.
In this case, FCOs corresponding to the cut-off magnitude, 34.9 mag (F775W), are located at $z_{max}$, which provides the luminosity of FCOs.
The number of FCOs was doubled to include faded FCOs, assuming the spatial density of the sum of FCOs and faded FCOs is constant.
Figure 16 shows $z_{max}$ dependence of the luminosity and mass for one FCO, and the mass-to-luminosity ratio.
The maximum mass-to-luminosity ratio is 0.79 at $z_{max}$=0.7, which is significantly lower than that of ordinary dwarf galaxies; thus,
the energy source of FCOs cannot be of a stellar origin. 

The two models described above are both overly simplified; thus, actual distribution may fall between the
two corresponding predictions. However, the
assumption that FCOs consist of missing baryons is confirmed in both models. FCOs contain half of the baryon mass
and are major constituents of the universe.

\begin{figure}
 \begin{center}
  \includegraphics[width=12cm]{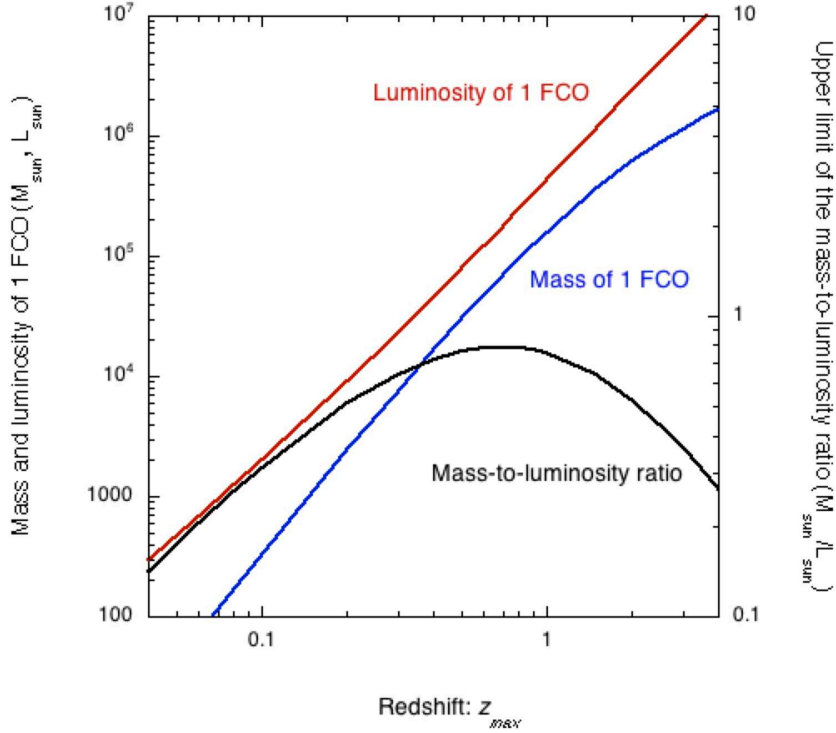} 
\end{center}
\caption{ Dependence of the  luminosity and mass of one FCO (left ordinate) and the mass-to-luminosity ratio (right ordinate) on 
the redshift, $z_{max}$, based on model 2. }
\label{fig16}
\end{figure}

\section{Discussion}
$\gamma$ ray observations are effective for studying the EBL because $\gamma$ rays are absorbed by collisions with
background photons. \citet{Biteau15} carefully analyzed all previous data of the TeV $\gamma$ observations with 
redshifts lower than 0.6, and concluded that the obtained EBL is basically consistent with the integrated light of the known galaxies. 
However, they found an excess EBL at optical wavelengths. 

\citet{Fermi18} searched for the EBL with lower energy $\gamma$ rays, and found that
the obtained EBL is consistent with evolution models of galaxies with a redshift range of 0.2 to 3. 

These $\gamma$ ray observations imply that FCOs must be low-redshift objects, most likely with redshifts lower than 0.1, if
FCOs are the source of the excess EBL. Since model 1
is not applicable for such a low redshift, we estimated the properties of FCOs with model 2.
For $z_{max}$=0.1, the luminosity and mass of FCOs correspond to $1.0 \times 10^{3}$ L$_{\odot}$ and $3.2 \times 10^{2}$ M$_{\odot}$, respectively,
and the mass-to-luminosity ratio is 0.32 M$_{\odot}$/L$_{\odot}$.
The comoving distance for $z_{max}$=0.1, is 421 Mpc, and the
spatial number density, including faded FCOs, reaches $8.8 \times 10^{6}$  (Mpc)$^{-3}$.
The distance to bright FCOs ($\sim $30 mag) for which we measured the radial profiles is 48 Mpc, and the
upper limit of the effective radius of 0.02 arcsec corresponds to 4.7 pc. 
These results indicate that the energy source of FCOs cannot be of a stellar or nuclear origin and that FCOs are much more compact
than an ordinary stellar system with the same mass. 

 Another important feature is the good correlation between the near-infrared background and the X-ray background.
\citet{Cappelluti13} and \citet{Cappelluti17} detected a good correlation of the background fluctuation between Spitzer 3.6 and 4.5 $\mu$m, and 
Chandra soft X-ray bands, [0.5-2] keV.  The correlation with harder X-ray bands, [2-7] keV, is marginally significant. 
The authors showed that the detected correlation at an angular scale of smaller than 20 arcsec can be attributed to 
known sources, such as AGNs, starburst galaxies, and hot gas, but residual fluctuation remains at large angular scales.
Since the X-ray background has a peak near 30 keV, the X-ray sources responsible for the residual fluctuation 
may be new unknown objects.
\citet{Cappelluti17} suggested accreting early black holes, including both DCBHs and primordial black holes.

\citet{Cappelluti17} found that the correlated X-ray background is less than $7\times 10^{-13}$ erg cm$^{-2}$ deg$^{-2}$, or
$2.3\times 10^{-3}$ nW m$^{-2}$ sr$^{-1}$, which corresponds to $\sim 10^{-4}$ of the lower limit of the excess optical background.
The ratio is very low compared with the ordinary quasar whose X-ray emission is $\sim$10\% of the optical emission \citep{Elvis94}.
This suggests that the X-ray emission mechanism of FCOs is not so efficient compared to that of the quasars.

\citet{M-W16} confirmed the correlation between the Spitzer 3.6 and 4.5 $\mu$m bands, and Chandra soft X-ray bands at small 
angular scales, but the fluctuation detected at angular scales larger than 20 arcsec was not significant.
The authors performed an additional correlation study with HST optical and infrared bands ($0.6 \sim 1.6$ $\mu$m), however, they did not obtain 
a significant correlation because the errors were too large.

Although correlations between the HST optical bands and the Spitzer 3.6 and 4.5 $\mu$m bands have not been directly confirmed, we conclude
that a certain level of correlation exists between optical and soft X-ray background, since various combinations of wavelength bands 
show a good cross-correlation for the wavelength range from the optical to Spitzer bands (\citet{Matsumoto05}, \citet{Matsumoto11}, \citet{Zemcov14}, \citet{M-W15}). 

These features are very similar to those of quasars. FCOs could be powered by gravitational energy associated with a 
black hole. It may be plausible to regard FCOs as mini-quasars, since the mass and luminosity scales of FCOs are much 
lower than those of quasars. However, the corresponding 
emission mechanism, particularly for the infrared excess, is not clear. 

FCOs produce not only optical and infrared background, but also X-ray background, and are very populous.
Even in our vicinity, there exist many faded FCOs, which may originate from gravitational waves \citep{Abbott16}.

FCOs appear to be major constituents of the universe, however, more observational data are needed to
delineate the origin of FCOs. 
In particular, spectroscopic observations and deeper surveys with better spatial resolution are needed. 
Observations toward the different sky regions are also important to examine the effect of the cosmic variance.
JWST will play a crucial role in these observations.

\section{Summary}
We performed a fluctuation analysis for the deepest part of the XDF using four optical bands with the following results.
\begin{itemize}
  \item Large spatial fluctuations which cannot be reproduced by ordinary galaxies, are detected for all bands.
  \item Good cross-correlations between wavelength bands are detected. In particular, an almost perfect cross-correlation is
 found between the F775W and F850LP bands.
  \item  The cross-correlation analysis results in a rather flat auto-correlation spectra down to 0.2 arcsec for the F775W and F850LP bands.
The fluctuation level provides a lower limit of 24 nW m$^{-2}$ sr$^{-1}$ for the absolute sky brightness, which is significantly
brighter than the ILG and is consistent with previous
HST and CIBER observations.
\end{itemize}

As a candidate object responsible for the excess brightness, we identified FCOs whose surface number density
rapidly increases towards the faint end. FCOs whose stellarities are larger than 0.9 have the following characteristics.  
 \begin{itemize}
   \item The dependence of the surface number density on the F775W band magnitude is found to follow $ \propto 10^{1.2m}$.
   \item The radial profiles of FCOs are indistinguishable from the PSF. Applying de Vaucouleur's law, we found an effective radius $r_{e}$
   of less than 0.02 arcsec.
   \item The SEDs of FCOs in the optical bands follow a power law, $\lambda \cdot I_{\lambda} \propto \lambda^{-1}$,
   while the SEDs of the infrared bands show an excess over the optical power law with structure.
   \item Presuming that FCOs are responsible for the excess background brightness, a cut-off magnitude
   must reach 34.9 mag for the F775W band. In this case, the surface number density of FCOs 
   is $3.35 \times 10^{10}$ (degree)$^{-2}$.   
\end{itemize}
   
$\gamma$ ray observations require that the excess background brightness have an origin with a redshift below 0.1.
This requirement provides restrictions for FCOs, and a simple model renders the following properties for FCOs.

\begin{itemize}
  \item The luminosity and mass of one FCO are 1.0  $\times$$10^{3}$  and 3.2  $\times$ 10$^{2}$ solar units, respectively.
  \item FCOs can account for missing baryons. 
  \item   The very low mass-to-luminosity ratio implies that 
           the energy source could be powered by gravitational energy associated with a 
           black hole. 
  \item The physical size of FCOs is less than 4.7 pc.
\end{itemize}
 
Considering the good correlation between the near-infrared and X-ray background, these results suggest 
that FCOs may be mini-quasars.

\begin{ack}
We thank T. Nakagawa, T. Yamada, T. Wada, N. Arimoto M., Onodera and M. Ouchi for the constructive discussions and variable comments. 
We are also grateful to anonymous referee for his/her valuable comments. T. M. would like to thank
KASI and ASIAA for providing a good research environment for this work.
This work was supported by JSPS KAKENHI Grant Number 26800112, 17K18789 and 18KK0089.
\end{ack}

\appendix

\section{Supplementary information on the masking}
We examined the dependence of the fluctuation spectra on the depth of the mask level to confirm the validity of 
the masking process. Figure 17 shows fluctuation spectra at the F775W band after the step 2, 3, and 4 (final). The
recovery factors are not adapted to avoid the large errors. Percentages of the remaining pixels are 74.3, 63.5, and 
53.3 \%, respectively. Considering the recovery factor, fluctuation after the step 3 is very close to 
that after the step 4, which indicates that extended wings are well masked after the step 4.
 
 \begin{figure}
 \begin{center}
  \includegraphics[width=12cm]{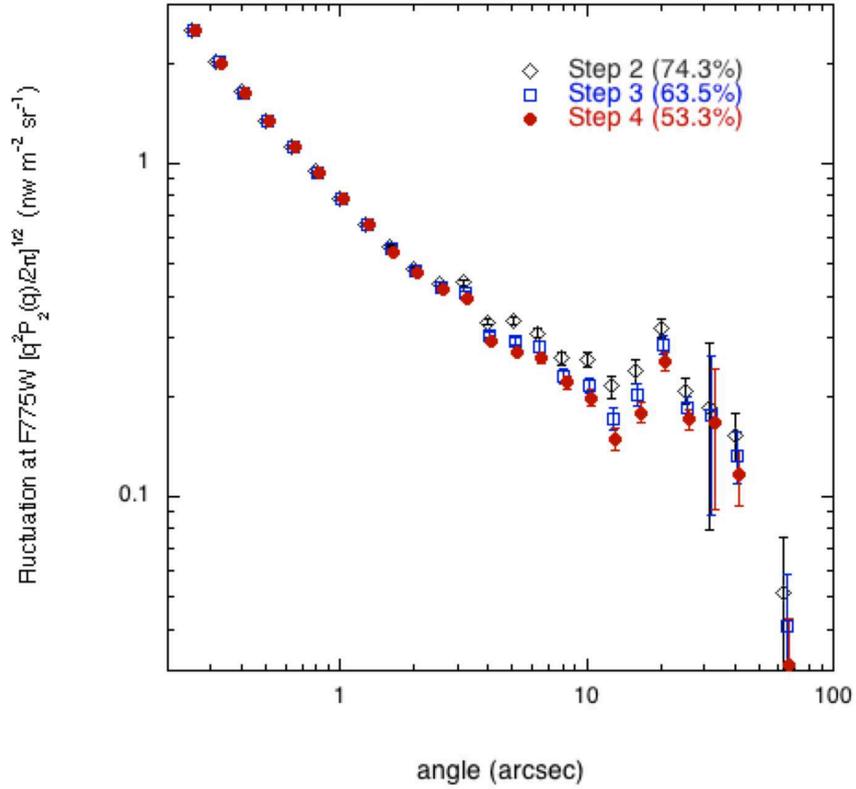} 
\end{center}
\caption{Dependence of the fluctuation spectra on the depth of the mask level (see text). Horizontal scales are a little shifted
to avoid ovelap of signals}
   \label{fig17}
\end{figure}

Figure 18 shows the histogram on the distribution of the surface brightness for $2 \times 2$ pixels resolution.
Observed distributions are well fitted by the Gaussian distribution, which is already known in the previous fluctuation
analyses (\citet{Thompson07}, \citet{Donnerstein15}).

 \begin{figure}
 \begin{center}
  \includegraphics[width=16cm]{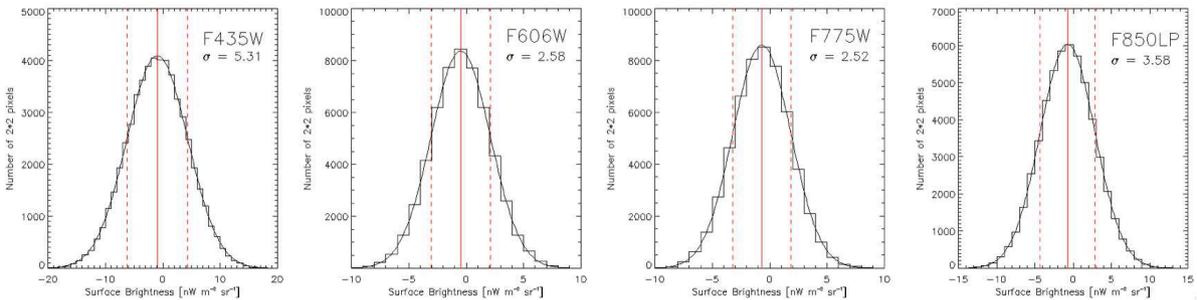} 
\end{center}
\caption{Histogram on the distribution of the surface brightness for $2 \times 2$ pixels resolution. 
The black solid lines indicate the result of the Gaussian fits. The red vertical solid and dotted lines show the average
values and $1 \sigma$ dispersion levels.}
   \label{fig18}
\end{figure}

Figure 19 shows the recovery factor which is obtained by comparing the fluctuations before and after masking. 
Errors are estimated after 100 times simulations. 

 \begin{figure}
 \begin{center}
  \includegraphics[width=12cm]{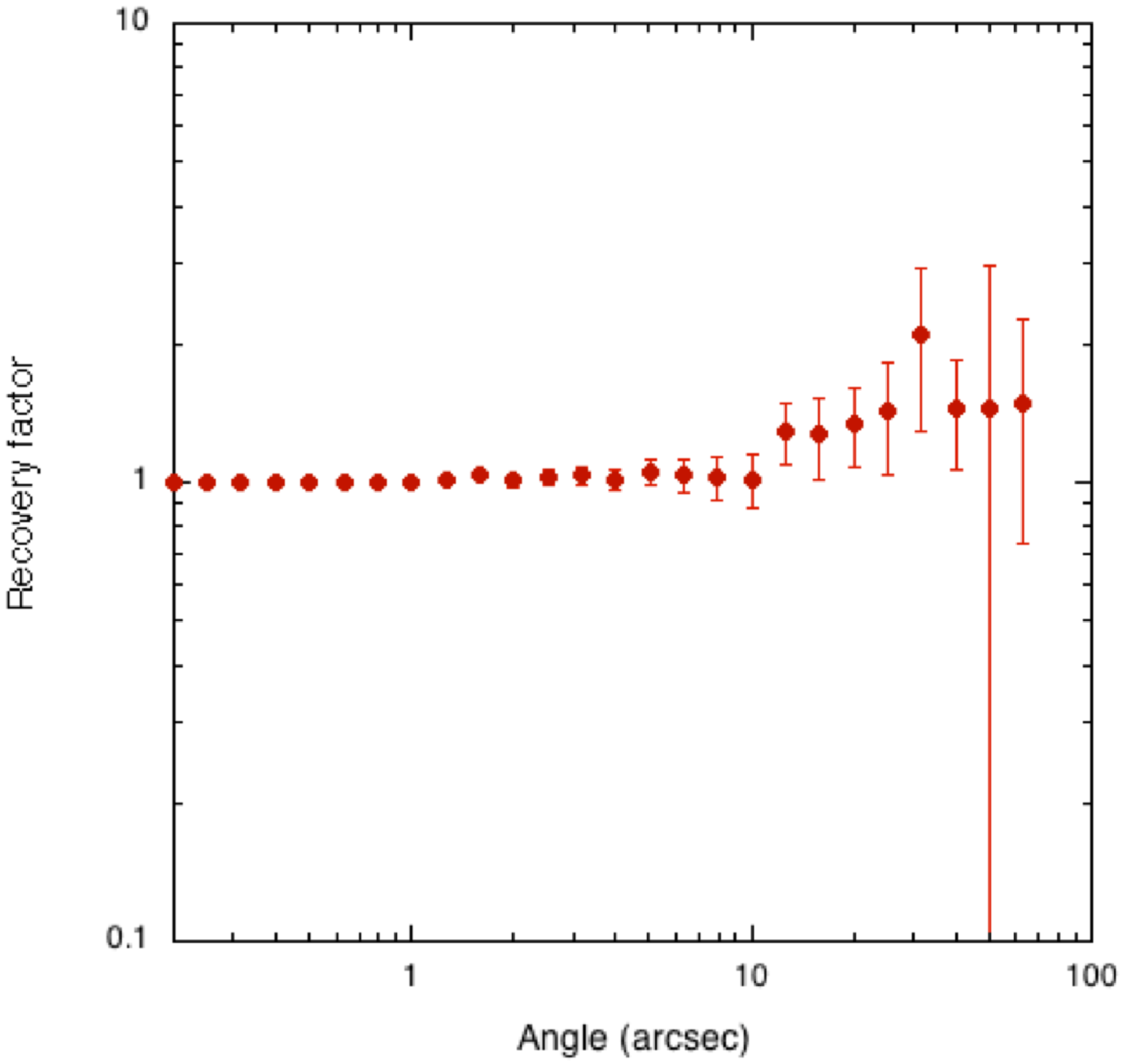} 
\end{center}
\caption{Recovery factor }
   \label{fig19}
\end{figure}

\section{Radial profiles and SEDs for 48 bright FCOs}
Figures 17 and 18 present radial profiles and SEDs for bright FCOs, respectively. 
Radial profiles are obtained for the combined F606W and F775W images, since the signal-to-noise ratios for the F435W 
and F850LP bands are relatively poor.
For bright sources, we excluded UDF2710, 6445, and 7307, since the SEDs based on the UDF catalog are substantially
different from those based on the UVUDF catalog. This difference is most likely due to source confusion because 
their coordinate data are also significantly different. 
Although the coordinate data for \#36 (UDF1240) are consistent, a discrepancy in the SED and a deviation of the radial profile from the 
PSF are found for this object, which may also be ascribed to source confusion, as there exists another 
source in the immediate vicinity.

\begin{figure}
 \begin{center}
  \includegraphics[width=16cm]{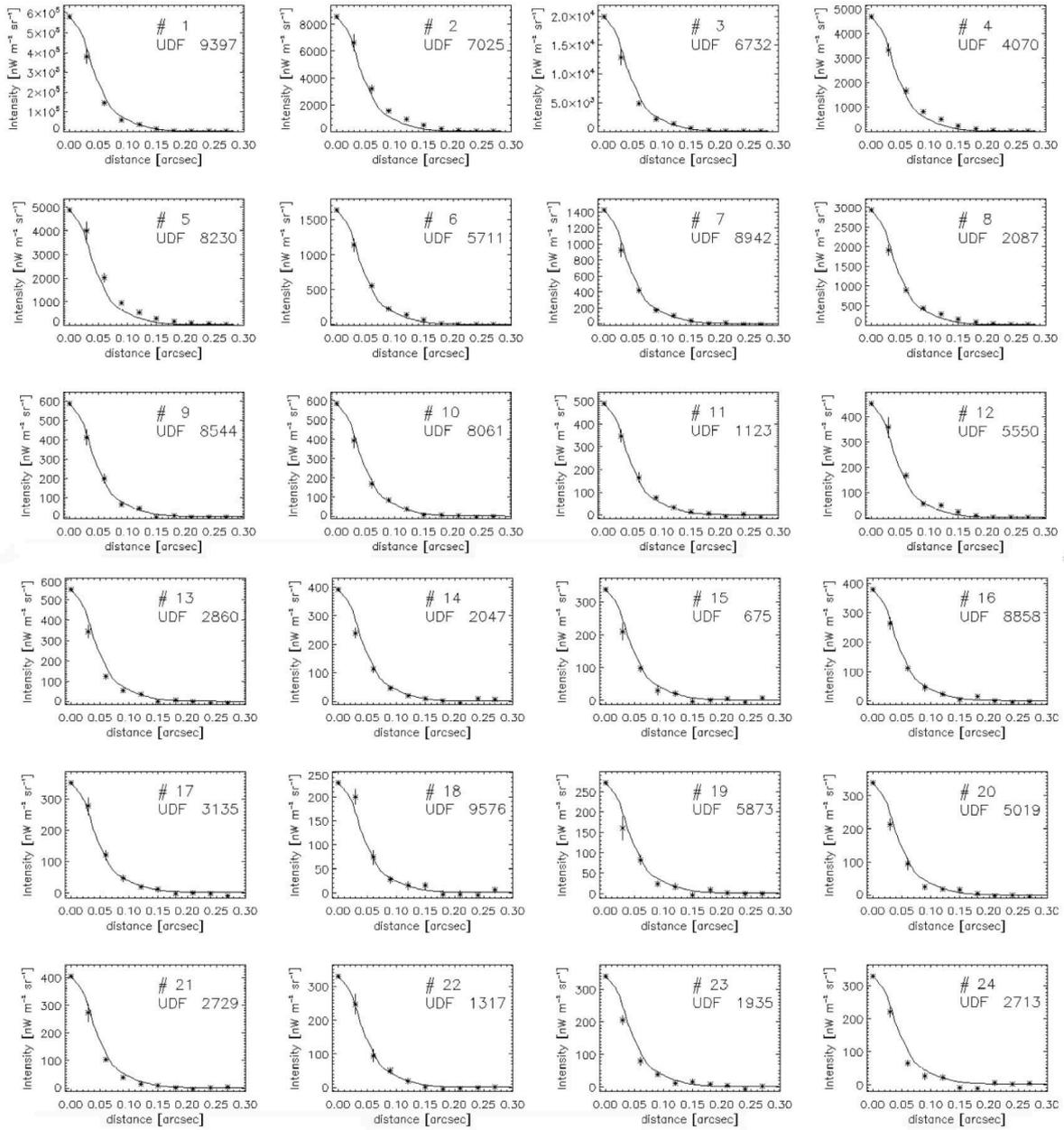} 
\end{center}
\caption{Radial profiles of bright FCOs for the combined images of F775W and F850LP. The solid lines present the same radial profiles
for the PSF obtained from stars. }
   \label{fig201}
\end{figure}

\renewcommand{\thefigure}{20}
\addtocounter{figure}{-1}

\begin{figure}
 \begin{center}
  \includegraphics[width=16cm]{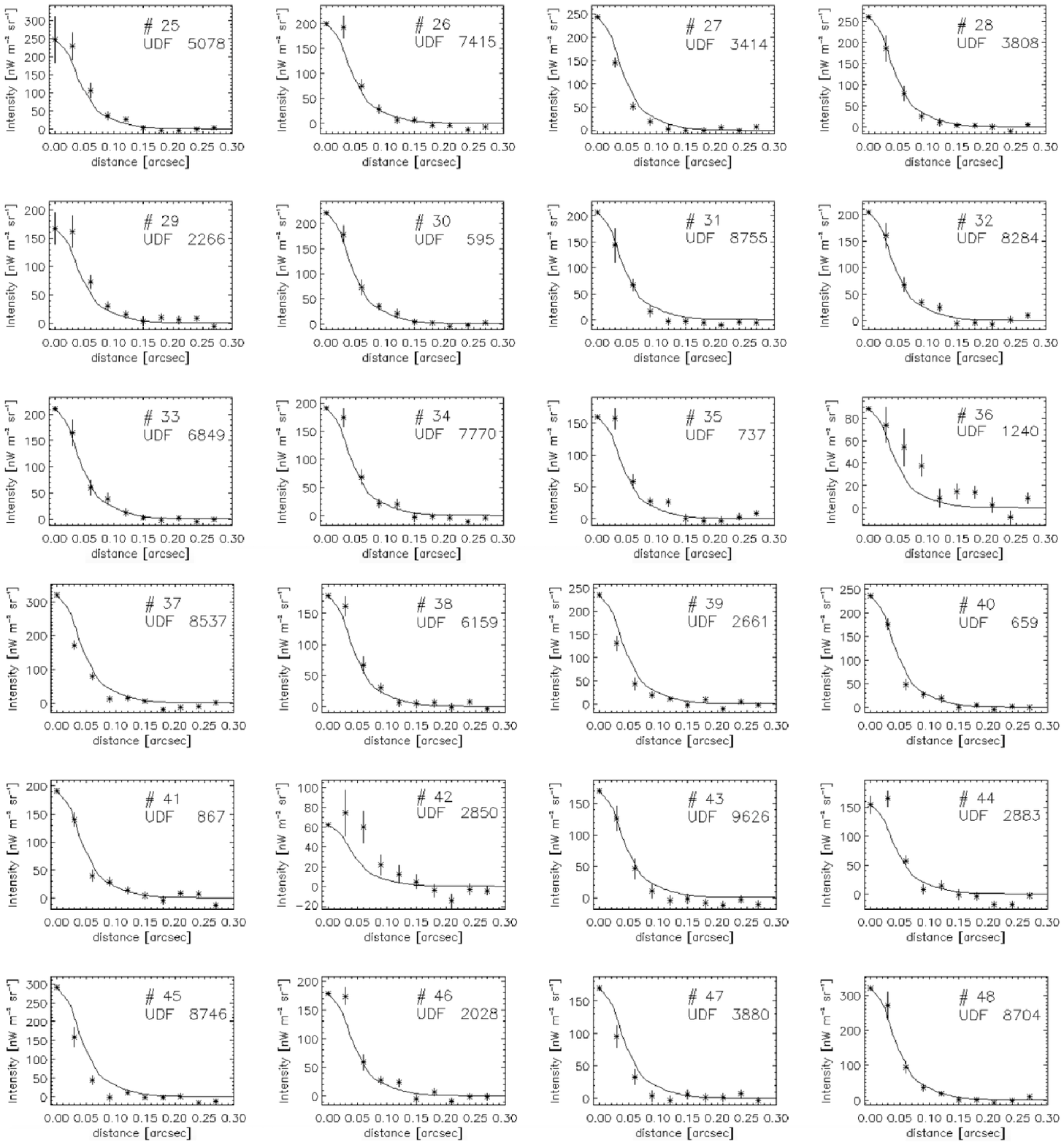} 
\end{center}
\caption{continued}
   \label{fig202}
\end{figure}

\renewcommand{\thefigure}{21}

\begin{figure}
 \begin{center}
  \includegraphics[width=16cm]{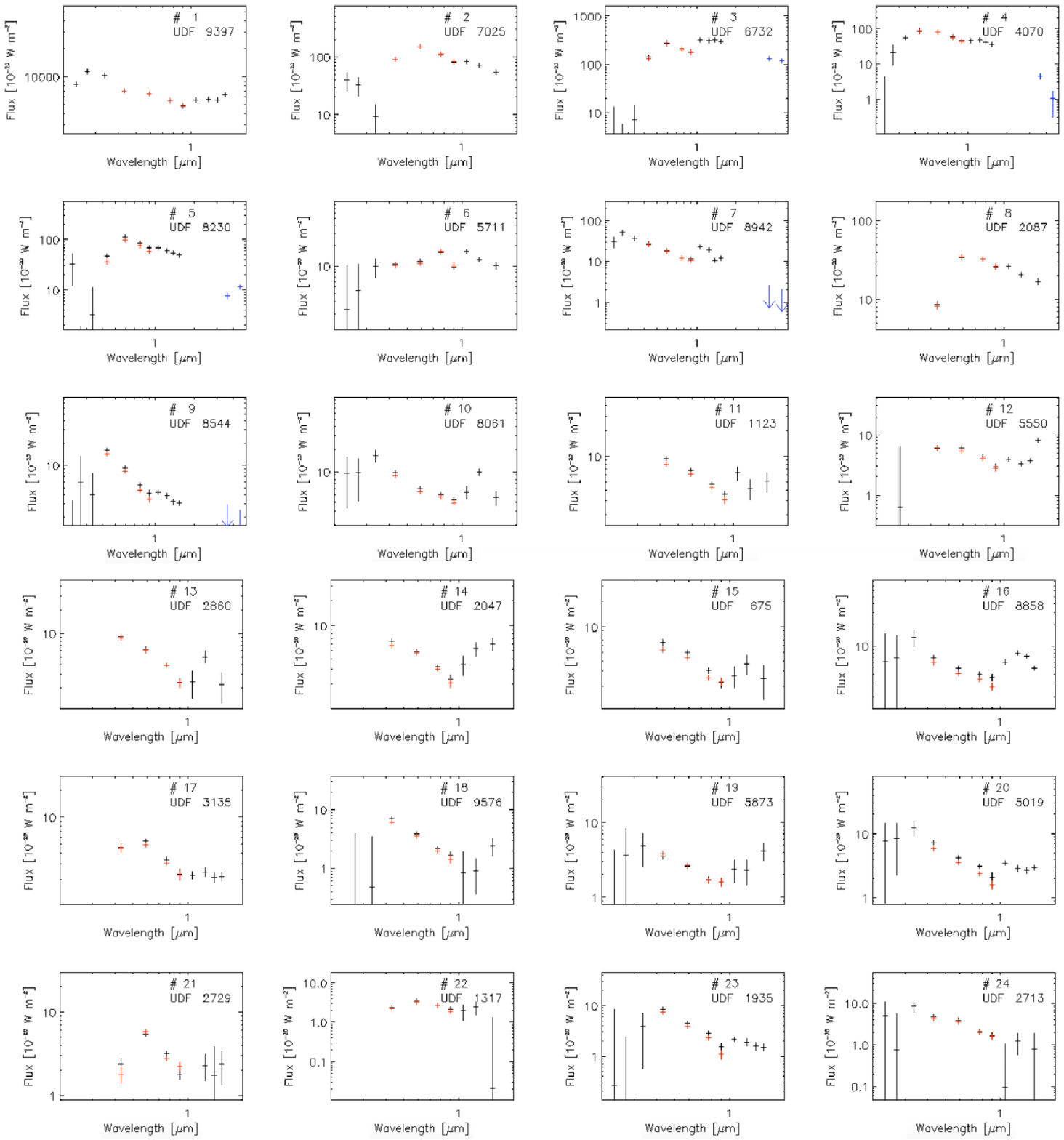} 
\end{center}
\caption{SEDs of 48 bright FCOs. The vertical bars indicate the errors. The black and red symbols denote the flux values
reported by \citet{Rafelski15} 
and \citet{Beckwith06}, respectively, while the blue symbols present those of the Spitzer bands as measured in this work.}
   \label{fig211}
\end{figure}

\renewcommand{\thefigure}{21}
\addtocounter{figure}{-1}

\begin{figure}
 \begin{center}
  \includegraphics[width=16cm]{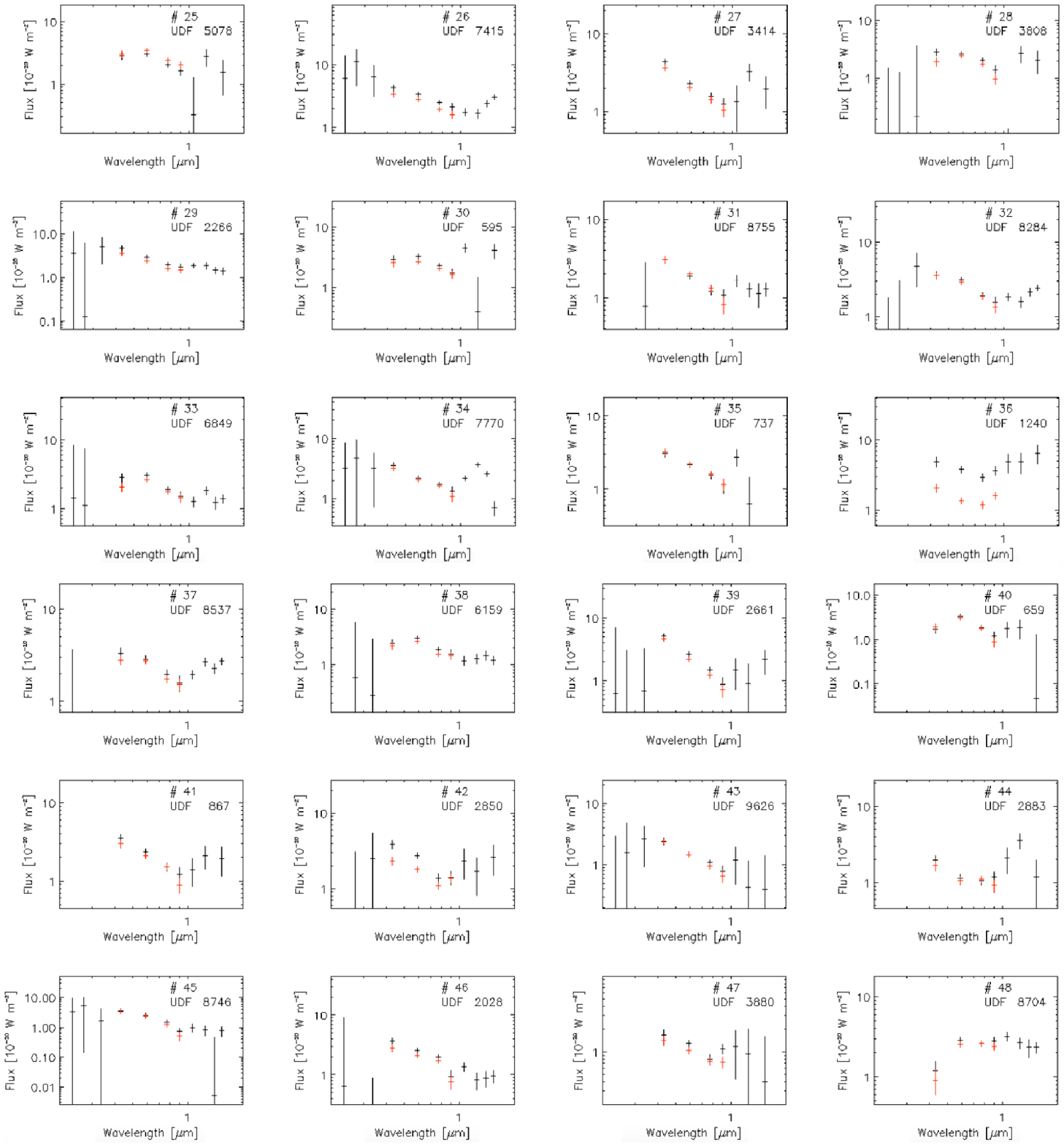} 
\end{center}
\caption{continued,}
   \label{fig212}
\end{figure}

\newpage

\end{document}